\documentclass[]{IEEEtran}

\usepackage{amsfonts}
\usepackage{tabu}
\usepackage{booktabs}
\usepackage{float}
\usepackage{gensymb}
\usepackage{textcomp}
\usepackage{dblfloatfix}
\usepackage{amsmath}
\usepackage{algorithm}
\usepackage{algpseudocode}
\usepackage{pifont}
\usepackage{verbatim}
\usepackage{float}
\usepackage{graphicx}
\PassOptionsToPackage{hyphens}{url}
\usepackage{hyperref}
\usepackage{seqsplit}
\usepackage{caption}
\usepackage{enumitem}
\usepackage{array}
\usepackage{subcaption}
\usepackage{booktabs}
\usepackage{breqn}
\usepackage{amsmath}

\usepackage{array}
\newcolumntype{A}[1]{>{\centering\let\newline\\\arraybackslash\hspace{0pt}}m{#1}}

\begin{document}


\title{Towards Inferring Mechanical Lock Combinations using Wrist-Wearables as a Side-Channel\thanks{This paper was revised on September 26, 2018. Please refer to the Appendix for details on corrections made.}}

\author{
\IEEEauthorblockN{Anindya Maiti\textsuperscript{$\dagger$}, Ryan Heard\textsuperscript{$\dagger$}, Mohd Sabra\textsuperscript{$\dagger$}, and Murtuza Jadliwala\textsuperscript{$\ddagger$}}\\
\IEEEauthorblockA{\textsuperscript{$\dagger$}Wichita State University\\
\textsuperscript{$\ddagger$}University of Texas at San Antonio\\
Email: \{axmaiti, rwheard, masabra\}@shockers.wichita.edu, murtuza.jadliwala@utsa.edu} \\

}

\maketitle

\begin{abstract}
Wrist-wearables such as smartwatches and fitness bands are equipped with a variety of high-precision sensors that support novel contextual and activity-based applications. The presence of a diverse set of on-board sensors, however, also expose an additional attack surface which, if not adequately protected, could be potentially exploited to leak private user information. In this work, we investigate the feasibility of a new attack that takes advantage of a wrist-wearable's motion sensors to infer input on mechanical devices typically used to secure physical access, for example, combination locks. We outline an inference framework that attempts to infer a lock's unlock combination from the wrist motion captured by a smartwatch's gyroscope sensor, and uses a probabilistic model to produce a ranked list of likely unlock combinations. We conduct a thorough empirical evaluation of the proposed framework by employing unlocking-related motion data collected from human subject participants in a variety of controlled and realistic settings. Evaluation results from these experiments demonstrate that motion data from wrist-wearables can be effectively employed as a side-channel to significantly reduce the unlock combination search-space of commonly found combination locks, thus compromising the physical security provided by these locks.
\end{abstract}


\IEEEpeerreviewmaketitle

\section{Introduction}
\label{intro}

Wrist-wearables such as smartwatches and fitness bands are gaining popularity among mobile users, and will continue to be a prevalent mobile technology in the future \cite{noauthor_wearables_nodate}. The presence of a diverse set of sensors on-board these devices, however, expose an additional attack surface which, if not adequately protected, could be potentially exploited to leak private user information. Weak or absent access control and security policies vis-\'{a}-vis some of these sensors have further compounded this problem. The research literature is rife with proposals that demonstrate how data from wrist-wearable sensors can be abused to infer private user information, such as, keystrokes, activities and behavior  \cite{maiti2015,wang2015,liu2015good,maiti2016smartwatch,wang2016friend,shoaib2016complex,xu2015finger,wen2016serendipity,karatas2016leveraging}. 
The continuous placement of wrist-wearables on users' body, coupled with their unique design and usage, also puts them at a significantly higher risk of being targeted for such privacy threats.

Our focus is on threats that enable an adversary to infer private inputs or interactions made by a target user on an input-interface (of some system of interest to the adversary) by taking advantage of \emph{zero-permission sensor} data available from the user's wrist-wearable. Zero-permission sensors (i.e., sensors that are not regulated by explicit user or system-defined access permissions) provide a relatively unobstructed attack surface to the adversary. A majority of research contributions in this direction have primarily focused on threats that attempt to infer private user inputs on interfaces of purely cyber or cyber-physical systems, for example, inference of keystrokes or taps on physical keyboards or touchscreen keypads  \cite{maiti2015,wang2015,liu2015good,maiti2016smartwatch,wang2016friend,holmes2016luxleak}. We focus on a slightly different kind of threat in this work which is to investigate the feasibility of inferring a target user's private inputs or interactions on the interface of a purely mechanical device by harnessing the sensor data available from the user's wrist-wearable. 
We specifically focus on inferring inputs on mechanical devices typically used to secure physical access (on doors and lockers), for example, \emph{combination locks}. 
Such privacy threats concerning mechanical safety devices, which may now be feasible due to the upcoming wearable device technology, has the potential of impacting the physical safety and security of users. 

Our specific research goal in this work is to investigate the feasibility of inferring unlock combinations of commercially-available mechanical combination locks and safes (Figure \ref{lockfigs}) by exploiting \emph{inertial} or \emph{motion} sensor data from wrist-wearables such as smartwatches. During the unlocking process of combination locks, the wrist on the unlocking hand undergoes perceptible and unique movements and rotations of its own, which is strongly correlated with the unlock combination. Our hypothesis is that, if these motions can be accurately captured and characterized, then it can be used to infer the lock's combination. Our objective is to validate the above hypothesis by empirically evaluating the accuracy and effort with which such an inference attack can be executed using wrist-wearables. In line with this objective, we make the following technical contributions:

\begin{enumerate}[leftmargin=*]
\item A novel \emph{motion-based combination key inference framework} comprising of: (i) an \emph{activity recognition component} for efficiently and accurately identifying unlocking-related data in the continuous motion data stream, (ii) a \emph{segmentation component} to separate and appropriately characterize motion data corresponding to each part of the multi-part combination or key, and (iii) an \emph{attack component} that maps the characterizations of the individual parts obtained from the previous steps to a (or a set of) valid combination key(s).
\item A \emph{comprehensive empirical evaluation} of the proposed attack framework in order to assess its performance: (i) on a commercially available padlock and safe, (ii) by using different key spaces, (iii) in a cross-device setting, (iv) in a cross-hand setting, and (iv) under real-life lock operation scenarios.
\end{enumerate}

\section{Related Work}
\label{related}

Threats that attempt to infer private information, user-contexts or user-activities by capturing related electromagnetic, acoustic, optical and/or mechanical emanations from a target device or user and employing them as information side-channels have been well-studied in the literature \cite{quisquater2001electromagnetic,kuhn2002optical,agrawal2003side,asonov2004keyboard,berger2006dictionary,backes2008compromising,backes2009tempest,vuagnoux2009compromising,backes2010acoustic,hayashi2014threat,Ali:2015:KRU:2789168.2790109,WANG201520,Li:2016:CMP:2976749.2978397}. 
With the advent of smartphones, researchers started focusing on employing the phone's on-board hardware and software sensors to investigate the feasibility of similar inference attacks \cite{uluagac2014sensory}. One notable sensor modality that now became available as an attack vector is the smartphone's inertial or motion sensors, such as, \emph{accelerometers} and \emph{gyroscopes}, which are capable of capturing fine-grained linear and angular motion of the user or object on which the phone was placed. Smartphone inertial sensors have been exploited to infer keystrokes on the phone itself as well as external keyboards \cite{barisani2009sniffing,marquardt2011sp,CaiC:2011,xu2012taplogger,OwusuHDPZ:2012}, to track user movements and locations \cite{han2012accomplice,ho2015pressure,narain2016inferring}, to infer private user activities \cite{ortizL:2015} and to decode human speech \cite{michalevsky2014gyrophone}. Similarly, smartphone microphone and/or magnetometer have also been exploited to infer private user information \cite{schlegel2011soundcomber} or trade secrets (such as 3D-printer designs) \cite{hojjati2016leave,song2016my,faruque2016acoustic}, private user activities \cite{rossi2013ambientsense} and natural handwriting \cite{Yu:2016:WAB}. Recently, aggregate power usage over a period of time available from the smartphone's power meter was used to track user movements and locations \cite{michalevsky2015powerspy}. 

The arrival of smartwatches and fitness bands have fueled a similar line of research in the area of private user-input, activity and context inference threats that take advantage of data available from sensors on-board these commercial wrist-wearable devices. However, unlike smartphones, as smart wearables are always carried by users on their body in the same natural position, the resulting continuous nature of sensor data available from them is more vulnerable to misuse and related inference threats more likely to succeed. Smartwatch motion sensors, similar to the smartphone case, have been exploited to infer keystrokes \cite{maiti2015,wang2015,liu2015good,maiti2016smartwatch,wang2016friend}, user-activities \cite{shoaib2016complex,liu2016complex}, handwriting \cite{xu2015finger,wen2016serendipity} and driving behavior \cite{karatas2016leveraging}. Recently, ambient light sensors on these devices have also been used to infer private keystroke information \cite{holmes2016luxleak}. Given this plethora of research results, it is clear that sensors on-board mobile and wearable devices pose a significant privacy threat. It is alarming though that common mobile and wearable device users are unaware of such threats \cite{CragerMJH:2017}. 

In this work, we
investigate the feasibility of a new kind of privacy threat, i.e., inferring unlock combinations of mechanical locks using wrist-wearable motion sensors, which has never been investigated before.
Several modern smart locks offer a numeric keypad which can be compromised using known smartwatch-based keystroke inference techniques in the literature \cite{maiti2015,wang2015,liu2015good,maiti2016smartwatch,wang2016friend}. However, in this work we target traditional rotation-based mechanical locks which are still very popular and where existing attack techniques will not work. Blaze \cite{blaze2004safecracking,blaze2003rights} systematically examined physical and design weaknesses in both combination and pin-tumbler locks. However, our primary contribution in this work is to show how external side-channel attacks can make even a securely designed lock vulnerable. 



\section{Adversary Model}
\label{adversary}

\begin{figure}[b]
\centering
\begin{subfigure}[b]{0.33\linewidth}
\includegraphics[width=\linewidth]{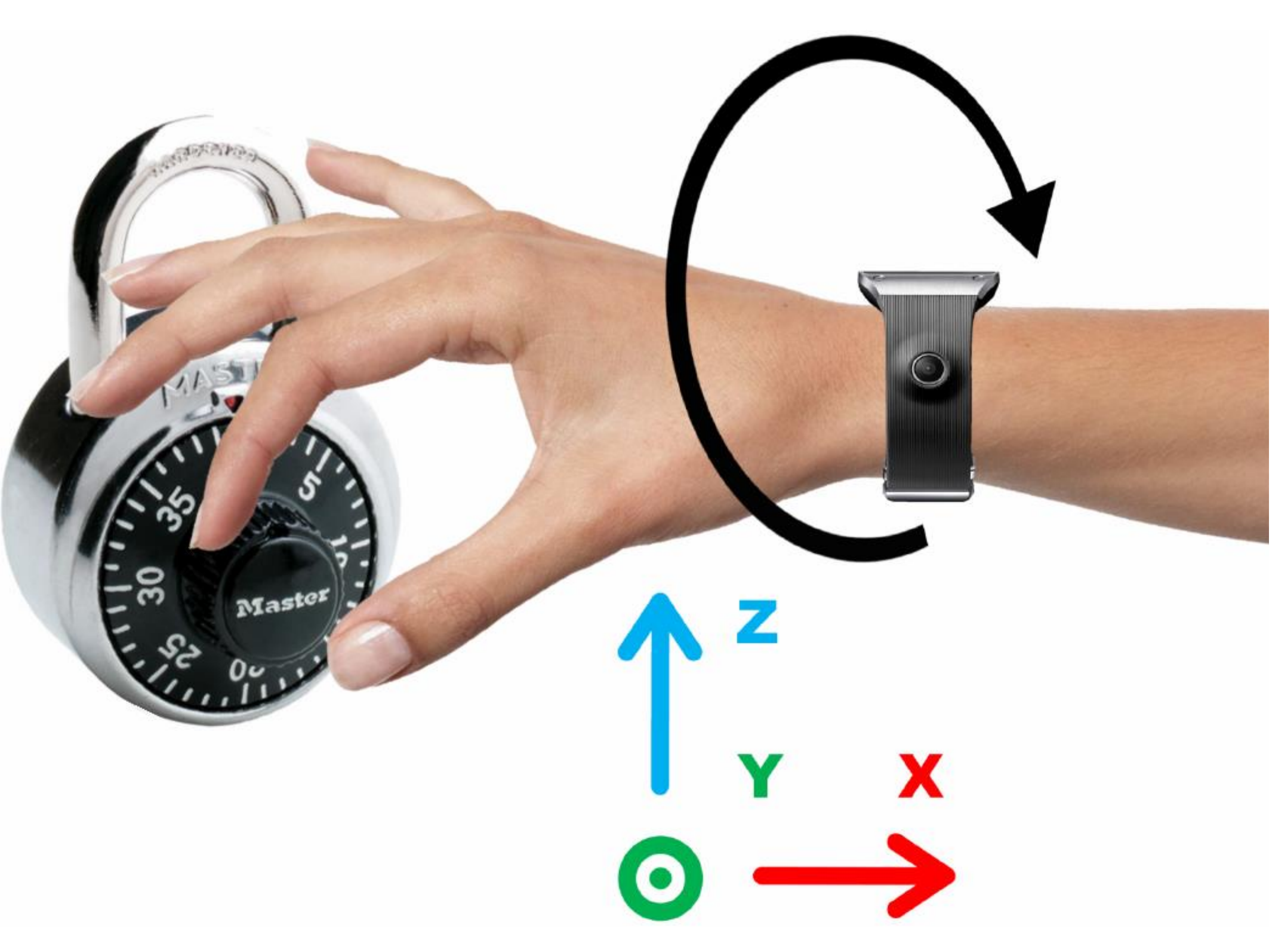}
\caption{}
\label{insidepadlocks}
\end{subfigure}
\quad\quad
\begin{subfigure}[b]{0.25\linewidth}
\includegraphics[width=\textwidth]{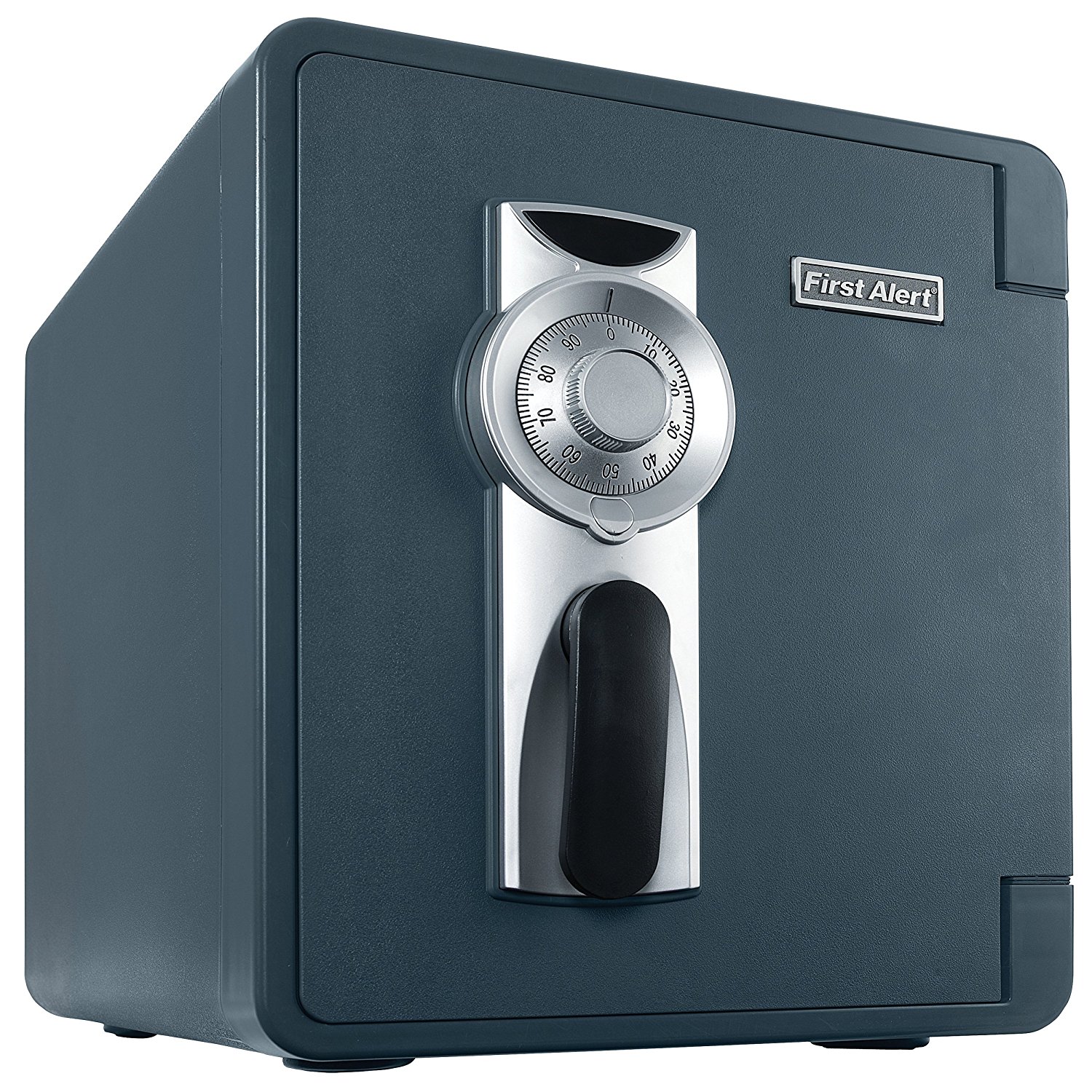}
\caption{}
\label{safefig}
\end{subfigure}
\caption{Targeted combination locks: (a) Master Lock 1500T padlock, (b) First Alert 2087F-BD safe.}
\label{lockfigs}
\end{figure}

We consider the scenario of a target user who is wearing a wrist-wearable such as a smartwatch and is entering the \emph{unlock combination} or \emph{key} on the circular dial of a mechanical combination lock (targeted by the adversary) with the watch-wearing hand. The goal of the adversary is to infer the unlocking combination of the lock by employing the inertial or motion sensor data available from the smartwatch worn by the target user. We assume that the adversary has knowledge of the exact type (make and model) of the target combination lock and that the dial of the lock has sufficient resistance to prevent rotation by mere movement of fingers.
The adversary is able to record and obtain the inertial or motion sensor data from the target smartwatch through several different modalities.
One way an adversary can achieve this is by creating a trojan app and then tricking the unsuspecting target user or victim into downloading and installing this trojan onto their wearable device. In case the adversary is a popular service provider, gaining access in such a fashion is much more straightforward as unsuspecting users may download and install the malicious app on their own volition. This malicious eavesdropping app samples the on-device sensors of interest (specifically, the \emph{gyroscope} sensor data is used for this particular attack) and transfers the sampled sensor data to a remote server controlled by the adversary through some covert communication channel, say by hiding it within useful communications.
We assume that the malicious app has the required permissions to access these sensors of interest. As the proposed attack employs the gyroscope sensor, which is a \emph{zero-permission} sensor on popular wearable operating systems such as Android Wear and watchOS, the adversary has a relatively unobstructed attack path once the malicious app is installed on the device. We also assume that the adversary maintains a remote server with sufficient storage and computational resources to archive the eavesdropped data and to perform \emph{offline inference computations}. The above adversary model is practically feasible and has been a standard assumption for similar lines of investigations. In addition to the above cyber resources, the adversary also has a limited amount of \emph{physical access} to the target lock (in order to conduct the actual physical attack on the lock by trying out the inferred combination), but not long enough to manually brute-force the lock's combination. The adversary presets/notes the position of the (lock's) dial before the target user begins the unlocking operation. However, the adversary has no visual access to the dial during the unlocking operation itself.

\section{Background}
\label{background}

\subsection{Mechanical Combination Locks}
\label{locks}
After studying the technical specifications of several commercially available mechanical combination locks, we decided to focus on two specific types of locks whose internal mechanical structure and physical operation are representative and commonly found in most rotary combination locks: (i) \emph{padlocks}, and (ii) \emph{consumer-grade safes}. For the padlock we chose a \emph{Master Lock 1500T} model lock (Figure \ref{insidepadlocks}), while for the safe we chose a \emph{First Alert 2087F-BD} safe (Figure \ref{safefig}). 

The front dial of the Master Lock 1500T is used to enter the unlock combination key and has 40 numbers on its face. As the combination key comprises of three numbers (each taking a value between 0 and 39) which must be entered sequentially, the resulting theoretical combination key space is  $40^3=64,000$. In order to unlock the Master Lock 1500T, a user must turn the dial clockwise two full rotations and stop at the first number of the combination key on the third turn (\emph{phase 1}), then turn it counter-clockwise past the first number of the combination key to the second number of the key (\emph{phase 2}), and finally turn the dial clockwise to the third number of the combination key (\emph{phase 3}). Let traversing from one number to it's sequential number (in any direction) be called a ``\emph{unit}" of traversal. Then it should be noted that, depending on the combination key being entered, in phase 1 the user traverses anywhere between 81 and 120 units in the clockwise direction, in phase 2 he traverses anywhere between 41 and 80 units in the counter-clockwise direction, and in phase 3 he traverses anywhere between 1 and 40 units in the clockwise direction. If this procedure is correctly followed, and if the entered combination key is correct, the indentations on the lock's cams align correctly allowing the hasp to be released and opening the lock. 

The First Alert 2087F-BD safe's lock dial comprises of 100 numbers (from 0 to 99) on its face. It's combination key comprises of four numbers (each taking a value between 0 and 99) which must be entered sequentially, thus resulting in a theoretical combination key space of $100^4$. In order to unlock the safe, a user must turn the dial counter-clockwise four full rotations and stop at the first number of the combination key on the fifth turn (\emph{phase 1}), then turn it clockwise twice past the first number to the second number (\emph{phase 2}), then turn it counter-clockwise past the second number to stop at the third number (\emph{phase 3}), and finally turn the dial clockwise to the fourth number (\emph{phase 4}). Depending on the combination key being entered, in phase 1 the user traverses anywhere between 401 and 500 units in the counter-clockwise direction, in phase 2 he traverses anywhere between 201 and 300 units in the clockwise direction, in phase 3 he traverses anywhere between 101 and 200 units in the counter-clockwise direction, and in phase 4 he traverses anywhere between 1 and 100 units in the clockwise direction. Similar to the Master Lock 1500T, if this procedure is correctly followed and if the entered combination key is correct, the safe opens.

\subsection{Combination Key and Wrist Movements}
\label{wristmovements}
Before designing an inference framework, we need to develop a clear understanding of how the activity of entering a combination key on a lock's dial impacts the wrist movement of the unlocking hand, and if it is possible to accurately and consistently characterize this movement using the motion sensor data obtained from modern wrist wearables such as smartwatches. More concretely, we would like to first understand the \emph{relationship between} the \emph{amount of movement of a lock's dial} and the corresponding \emph{amount of movement of the user's wrist}. We quantify the amount of movement of a lock's dial using the parameter \emph{transition}, which measures the \emph{number of units traversed when inputing a particular number of the combination}. As the unlock combination key of the Master Lock 1500T padlock has three numbers (and correspondingly, the unlocking procedure has three phases), the amount of movement of the lock's dial during the unlocking process can be completely characterized by three transitions. Similarly, as the First Alert 2087F-BD safe has a four number combination, the amount of movement of the lock's dial during unlocking can be completely characterized by four transitions. We quantify the amount of movement (or rotation) of a user's wrist by computing the \emph{angular displacement} from the observed \emph{smartwatch gyroscope} data. As the gyroscope measures angular velocity, the corresponding angular displacements can be calculated by integrating the obtained angular velocity readings.

\begin{figure}[b]
\centering
\begin{subfigure}[t]{0.49\linewidth}
\includegraphics[width=\textwidth]{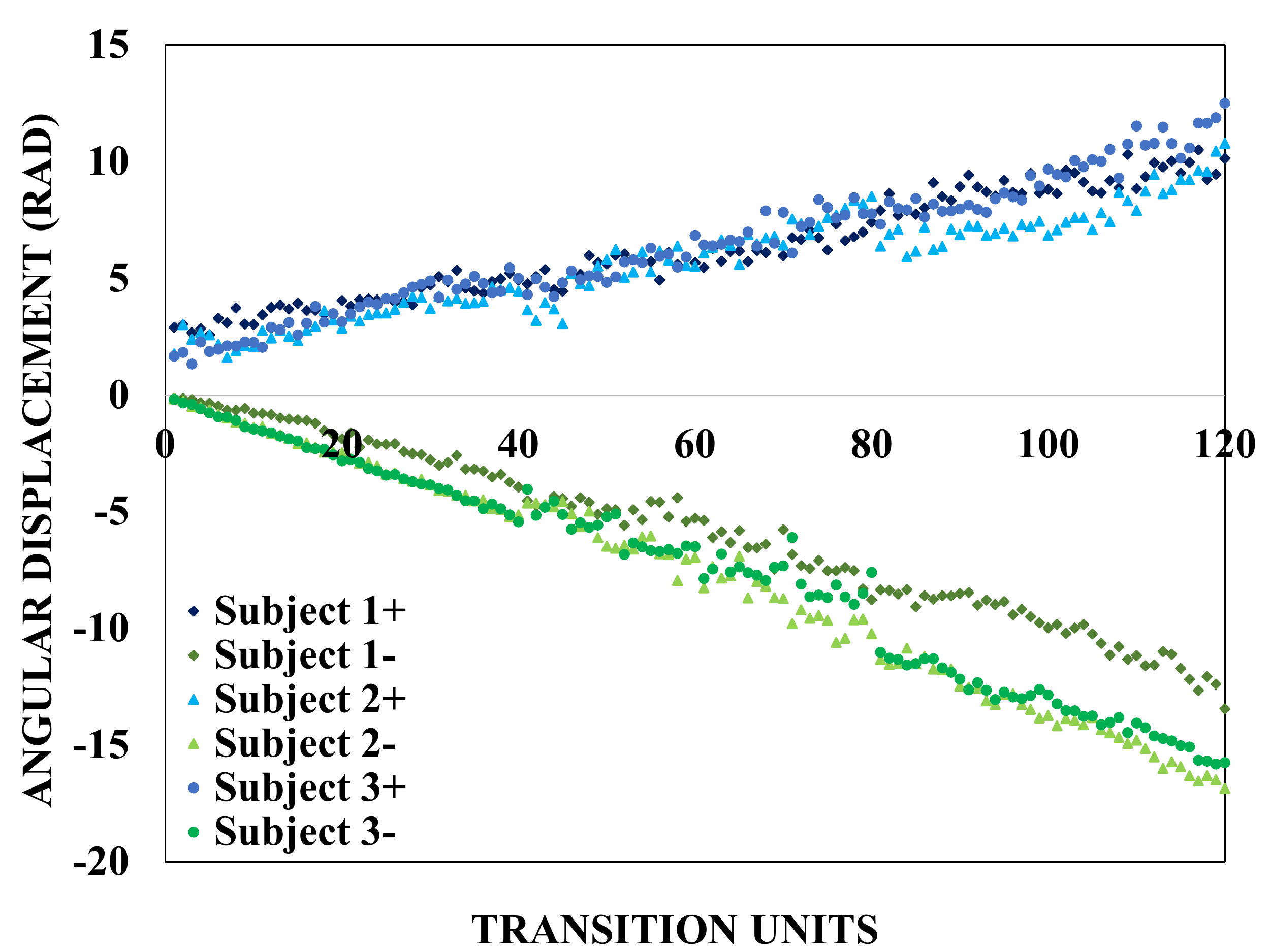}
\caption{}
\label{sub1a}
\end{subfigure}
\begin{subfigure}[t]{0.49\linewidth}
\includegraphics[width=\textwidth]{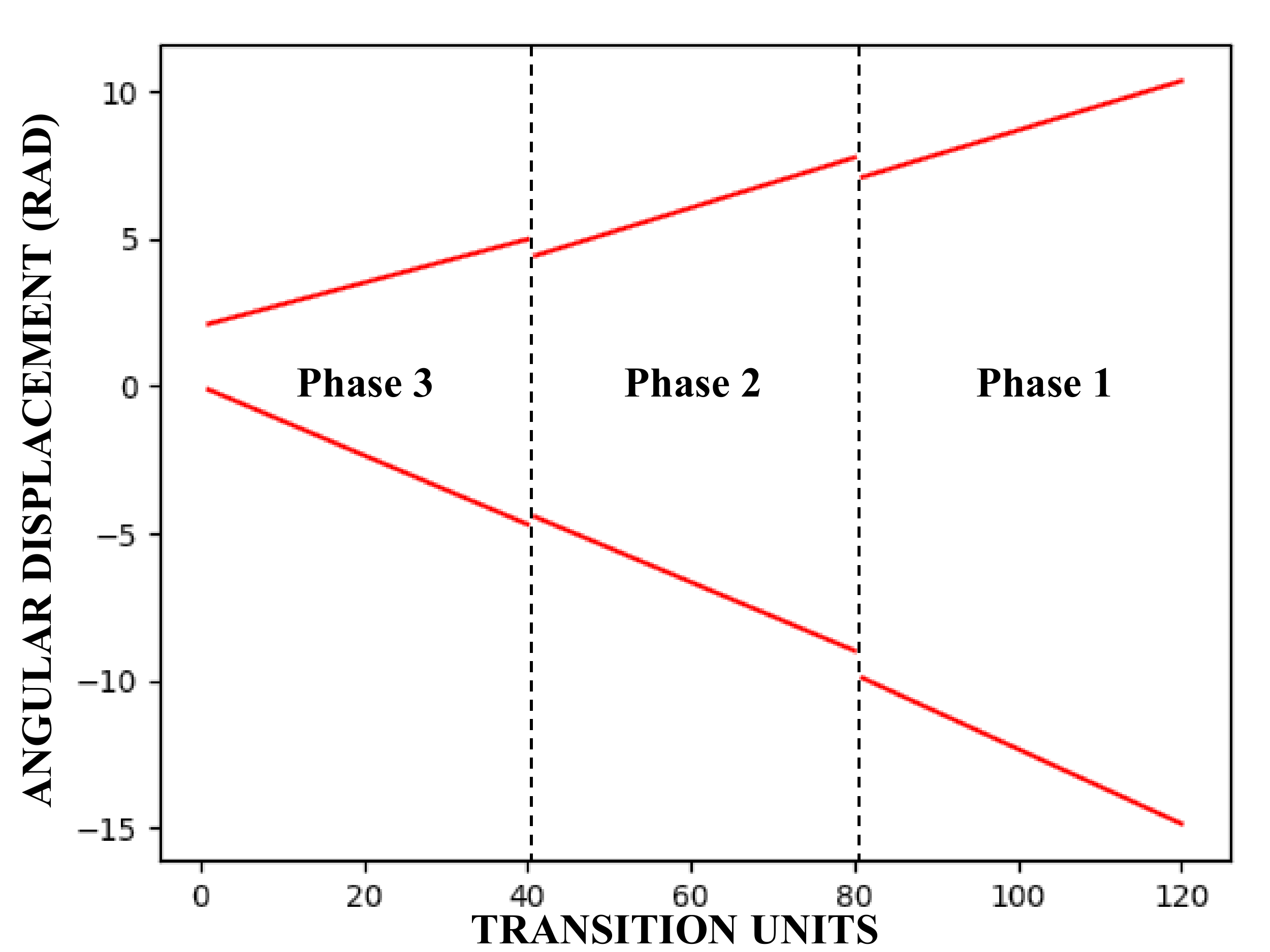}
\caption{}
\label{leastmeansq}
\end{subfigure}
\caption{(a) -- Positive (blues) and negative (greens) angular displacements, collected from three subjects; (b) Combined linear least squares fitting.}
\label{rotationgraph}
\end{figure}

In order to quantify the relationship between transitions on a lock's dial and the wrist's angular displacements, we conduct some preliminary unlocking experiments on the Master Lock 1500T padlock. Specifically, we collected smartwatch gyroscope samples at a sampling rate of 200 $Hz$ from three human subjects who unlocked the padlock wearing a Samsung Gear Live.
The subjects in our preliminary experiments entered 40 different combinations on the Master Lock 1500T padlock which covered all the 120 possible transitions (40 possible transitions per number in any combination key). While entering each combination, the subjects always started from a known position (number 0)\footnote{The starting point can be any number on the dial. However, the key inference function (Equations \ref{padlockkeyformula} and \ref{safekeyformula}) must be initialized accordingly, during the inference phase.} and entered the combination by correctly following the unlocking procedure described in Section \ref{locks}. For each subject, we plot the angular displacement (in radians), calculated by integrating the corresponding angular velocities observed on the $x$-axis of the smartwatch's gyroscope, for each each transition in either direction (Figure \ref{sub1a}).

From Figure \ref{rotationgraph} we first observe that, for each transition (irrespective of the direction of rotation), the angular displacement of the wrist calculated from the raw smartwatch gyroscope data is not the same as the angular displacement of the lock's dial. These inaccuracies could be attributed to the discrete nature of the gyroscope readings, which are limited by the maximum sampling rate of the gyroscope hardware. In addition to this, the cartilaginous joints between the fingers and the wrist, do not allow for a \emph{perfect rotation} of the wrist during the unlocking operation. As a result, an adversary cannot simply use the angular displacement of the wrist calculated from the raw smartwatch gyroscope data to determine the angular displacement, and thus the corresponding transition, on the lock's dial. Our second observation is that the angular displacement of the wrist calculated from the raw smartwatch gyroscope data can be approximated as an \emph{increasing linear function} of the transitions on the lock's dial. Although intuitive, the interesting and encouraging aspect here is that this relationship is consistent for all three subjects. Lastly, we observe that this linear relationship is reasonably \emph{homologous} or similar across different subjects. 
We only used the $x$-axis of the gyroscope data for these plots because we observed that the $x$-axis remains perpendicular to the lock (Figure \ref{insidepadlocks}) during the unlocking operation and provides a more accurate measure of angular displacement than the other two axes.

So, what do these observations mean to an adversary who wants to infer the combination key entered by some target user? The adversary is \emph{unable} to accurately determine the angular displacement or transition on the lock's dial (and thus the corresponding number in the combination) directly from the corresponding angular displacement of the wrist computed using the smartwatch's gyroscope data. However, an adversary could use the above observations to construct a \emph{learning-based inference framework} that translates angular displacements of the wrist (computed from the smartwatch's gyroscope data) to transitions on the lock's dial, and train this framework using some representative training data. The adversary could then employ such a trained inference framework to infer the combination (entered by the target user) from the smartwatch gyroscope data. We develop such an inference framework in Sections \ref{dattack} and \ref{pattack}. However, there are two additional challenges that we need to overcome. First, in a long sequence of time-series gyroscope data, how does the adversary \emph{identify} data corresponding to the unlocking motion? Second, to accurately compute the angular displacement of the wrist for each phase of the unlocking procedure, the adversary needs to \emph{divide} or \emph{segment} the gyroscope time-series into individual phases. We address these issues by developing an unlocking activity recognition technique (Section \ref{activityrecognition}), and a segmentation technique (Section \ref{segment}).

\subsection{Unlocking Activity Recognition}
\label{activityrecognition}
Before attempting to infer combinations from the target user's wrist motions, one critical challenge for the adversary is to precisely detect when the unlock event takes place. In order to overcome this challenge, we design an \emph{offline activity recognition technique} to detect and record timestamps of unlocking operations on combination locks. Our activity recognition technique does not require any additional adversarial capabilities or resources as it employs only the gyroscope data stream (specifically, the $x$-axis data) which is already recorded by the adversary for the inference task. While analyzing characteristics of the time-series gyroscope data during unlocking, we observed that the integrated angular displacement increases on both positive and negative axis in successive periods. This is because after rotating the dial (clockwise or counter-clockwise) to an extent, users release the dial, go back in reverse (counter-clockwise or clockwise, respectively), again grab the dial, and continue entering the remaining part of the combination key (clockwise or counter-clockwise, respectively). We refer to one such clockwise-counterclockwise (or vice-versa) motion during combination key entry as a \textit{``spin''}, which is primarily related to the comfortable wrist rotation ability (or desire) of humans. Such \textit{spin}-ing is repeated multiple times during any combination key entry, approximately every half a turn ($\pi$) and over a maximum duration of approximately 5 seconds. We can observe this phenomenon in the sample gyroscope ($x$-axis) time-series corresponding to a padlock unlocking operation (Figure \ref{splitting}). We utilize the above observations in the design of the following four features which will be employed by our activity recognition technique:

\begin{itemize}[leftmargin=*]
\item \textbf{Positive Displacements ($\textsuperscript{+}\alpha$):} Integration of positive $x$-axis gyroscope samples.
\item \textbf{Negative Displacements ($\textsuperscript{-}\alpha$):} Integration of negative $x$-axis gyroscope samples.
\item \textbf{Summed Displacement ($\textsuperscript{+}\alpha + \textsuperscript{-}\alpha$):} Sum of integrated positive and negative $x$-axis samples.
\item \textbf{Total Displacement Magnitude ($\textsuperscript{+}\alpha + |\textsuperscript{-}\alpha|$):} Sum of the magnitudes of integrated positive and negative $x$-axis samples.
\end{itemize}

In order to confirm the above observations, we computed the means and standard deviations of the above four features over all the 5 second windows (maximum duration of a \textit{spin}) in the preliminary unlocking-related gyroscope data collected earlier (Figure \ref{sub1a}). We observed that the mean values of the magnitudes of $\textsuperscript{+}\alpha$ and $\textsuperscript{-}\alpha$ are approximately similar in a \textit{spin}, the mean value of the total displacement magnitude is approximately double of both $\textsuperscript{+}\alpha$ and $\textsuperscript{-}\alpha$, and the mean value of the summed displacement is close to zero. 
We employ these learned mean and standard deviation values to form a \emph{decision-tree} for detecting \textit{spins}. During the activity recognition, the above four features are recursively computed for every 5 second window, and the decision-tree classifies a window as a \textit{spin} if all the four features are within one standard deviation of the learned means. 
In the case of padlock, if 5 (minimum number of \textit{spins} observed for the shortest padlock combination: 39-0-39) or more spins are observed within a short time window (empirically determined based on the maximum unlocking time observed in data) an unlocking activity is recognized. A similar strategy could be used to recognize unlocking operation on a safe.

\subsection{Segmentation}
\label{segment}

\begin{figure}[t]
\centering
\includegraphics[width=0.91\linewidth]{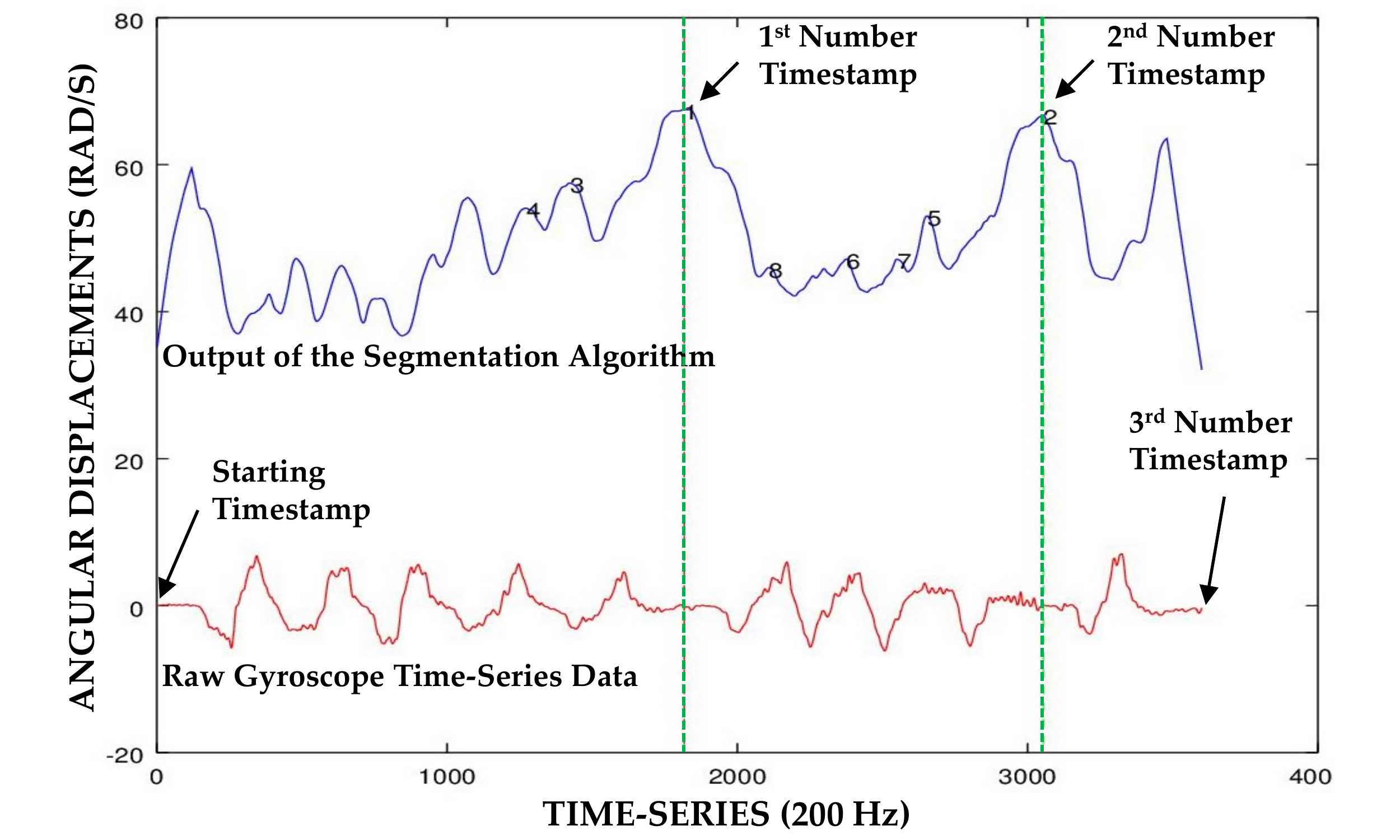}
\caption{Segmentation using a Gaussian filter.}
\label{splitting}
\end{figure}

Segmentation of the time-series gyroscope data representing the entire combination key input into data corresponding to individual phases or transitions (three for the padlock and four for the safe) will simplify the overall design of the inference framework. This is because the combination inference problem can then be reduced to the problem of independently inferring the combination number corresponding to each segmented transition. In order to design a reliable segmentation technique, we leverage on the observation from our earlier experiments that \textit{humans tend to slow down when they approach a number in their combination key}. We believe that this phenomenon is due to the cognitive processing of the human brain governing the physiological action of stopping at a particular number, which causes the subjects to slow down when approaching the intended number in their combination key or risk overshooting it (and thus having to restart the entire key entry process). We can observe this phenomenon in the time-series gyroscope data corresponding to the unlocking operation of the Master Lock 1500T padlock by one of the subjects, where we can clearly see (Figure \ref{splitting}) the sharp decreases in the angular velocity (red line) when approaching the combination key number near the end of each phase.
In order to \emph{automate} the process of segmentation, we design an algorithm to detect the relative decrease in angular velocity, and use the peaks (representing slowest movement) to segment the entire time-series. The algorithm first computes the absolute values of all samples in the gyroscope time-series data, inverts, and then amplifies the time-series by a factor of 10 (for better visualization). Then, on the resultant time-series, a Gaussian filter with a moving window \cite{deisenroth2011general} of 15 samples (learned empirically, at 200 Hz sampling frequency) is applied. Finally, the algorithm  performs a search for top-$2$ global peaks in the resultant time-series, which represents approximate timestamps for the first and second number of the combination key, in chronological order. The blue (top) line in Figure \ref{splitting} is an example of the visualized output of our segmentation algorithm, showing the detected peaks and resulting segmentation timestamps. Our algorithm also works on gyroscope data from the safe, using top-$3$ peaks.

\section{Deterministic Attack Framework}
\label{dattack}
We develop two learning-based inference frameworks to infer numbers of the combination key inputted on the lock's dial from the segmented smartwatch gyroscope data. We first outline a \emph{deterministic framework} which outputs a single inferred combination key from the segmented time-series gyroscope input. 

\subsection{Padlock Attack Model}
\label{detpadlock}
Assuming that the starting point $s$ on the padlock's dial is fixed/known (say, to be 0), we can define $\Phi_{1}=\{81,82..,120\}$, $\Phi_{2}=\{41,42..,80\}$ and $\Phi_{3}=\{1,2..,40\}$ as the sets of \emph{possible padlock transitions} in phase 1, phase 2 and phase 3 of the unlocking procedure, respectively. Now for a given 3-number combination key $k=\langle a,b,c\rangle$ of the Master Lock 1500T padlock, where $a,b,c \in \{0,1..,39\}$, let $\theta^{sa}_k\in \Phi_{1}$, $\theta^{ab}_k\in \Phi_{2}$ and $\theta^{bc}_k\in \Phi_{3}$ be the \emph{actual transitions} or number of units traversed (on the lock's dial) between consecutive numbers of the combination key $k$, i.e., $\theta^{sa}_k$, $\theta^{ab}_k$ and $\theta^{bc}_k$ are the number of units traversed between $0$ and $a$, between $a$ and $b$, and between $b$ and $c$, respectively. 
Let $\alpha^{sa}_k$, $\alpha^{ab}_k$ and $\alpha^{bc}_k$ denote the corresponding angular displacements of the target user's wrist (ignoring the direction or sign) calculated from the segmented smartwatch gyroscope data. The inference framework comprises of a \emph{training phase} and an \emph{attack phase}. During the training phase, the adversary collects training data (from a set of human participants) comprising of a set of $\theta$ and corresponding $\alpha$ values for a sample set of combinations covering all possible transitions. 
As indicated by our preliminary results (Figure \ref{sub1a}), the relationship between angular displacements of the wrist and transitions on the lock's dial can be approximated by a linear function. Thus, the adversary can use the training data to learn such a linear function $\alpha=m\theta+n$ that best fits all $(\theta$, $\alpha)$ points in each of the $[s,a]$, $[a,b]$ and $[b,c]$ transition ranges of the training data. 
The adversary can employ a \emph{least squares} \cite{york1966least} technique in order to learn such a linear function (Figure \ref{leastmeansq}). 
Then during the attack phase, for an unknown combination key $\hat{k}=\langle \hat{a},\hat{b},\hat{c}\rangle $, the adversary first segments the gyroscope data and computes the corresponding angular displacements $\alpha^{s\hat{a}}_{\hat{k}}$, $\alpha^{\hat{a}\hat{b}}_{\hat{k}}$ and $\alpha^{\hat{b}\hat{c}}_{\hat{k}}$. 
The adversary's goal then is to determine a combination $k'$, as an inference of $\hat{k}$, by first \emph{approximating} or \emph{estimating} the $\theta^{s\hat{a}}_{\hat{k}} \in \Phi_{1}$, $\theta^{\hat{a}\hat{b}}_{\hat{k}} \in \Phi_{2}$ and $\theta^{\hat{b}\hat{c}}_{\hat{k}} \in \Phi_{3}$ values from the corresponding angular displacements ($\alpha^{s\hat{a}}_{\hat{k}}$, $\alpha^{\hat{a}\hat{b}}_{\hat{k}}$ and $\alpha^{\hat{b}\hat{c}}_{\hat{k}}$, respectively). Let these approximations of $\theta^{s\hat{a}}_{\hat{k}}$, $\theta^{\hat{a}\hat{b}}_{\hat{k}}$ and $\theta^{\hat{b}\hat{c}}_{\hat{k}}$ be denoted as $\bar{\theta}^{s\hat{a}}$, $\bar{\theta}^{\hat{a}\hat{b}}$ and $\bar{\theta}^{\hat{b}\hat{c}}$, respectively. In order to accomplish this, the adversary employs the linear function ($\alpha=m\theta+n$) learned earlier. Once the transition in each phase has been estimated, $k'$ can be computed as:
\begin{equation}
\small
\begin{aligned}
k'=\langle ((-\bar{\theta}^{s\hat{a}}+s)\mod{40}), \\((\bar{\theta}^{\hat{a}\hat{b}}+(-\bar{\theta}^{s\hat{a}}+s))\mod{40}),\\ ((-\bar{\theta}^{\hat{b}\hat{c}}+(\bar{\theta}^{\hat{a}\hat{b}}+(-\bar{\theta}^{s\hat{a}}+s)))\mod{40})\rangle 
\end{aligned}
\label{padlockkeyformula}
\end{equation}

\subsection{Safe Attack Model}
\label{detsafe}
Similar to the padlock, we can define $\Psi_{1}=\{401,402..,500\}$, $\Psi_{2}=\{201,202..,300\}$, $\Psi_{3}=\{101,102..,200\}$ and $\Psi_{4}=\{1,2..,100\}$ as the sets of \emph{possible safe transitions} in phase 1, phase 2, phase 3 and phase 4 of the safe unlocking procedure, respectively.
For a given 4-number safe combination $k=\langle a,b,c,d\rangle $, where $a,b,c,d \in \{0,1..,99\}$, let $\theta^{sa}_k\in \Psi_{1}$, $\theta^{ab}_k\in \Psi_{2}$, $\theta^{bc}_k\in \Psi_{3}$ and $\theta^{cd}_k\in \Psi_{4}$ be the \emph{actual transitions} between consecutive numbers of the combination key $k$.
Also, let $\alpha^{sa}_k$, $\alpha^{ab}_k$, $\alpha^{bc}_k$ and $\alpha^{cd}_k$ denote the corresponding angular displacements of the target user's wrist (ignoring the direction) calculated from the segmented smartwatch gyroscope data.
Similar to the padlock case, the adversary collects training data (from a set of human participants) comprising of a set of $\theta$ and corresponding $\alpha$ values for a sample set of combinations covering all possible transitions, and uses it to learn a linear function (of the form of $\alpha=p\theta+q$) by employing a least squares \cite{york1966least} technique. Then during the attack phase, for an unknown combination key $\hat{k}=\langle \hat{a},\hat{b},\hat{c},\hat{d}\rangle $, the adversary first segments the time-series gyroscope data and computes the corresponding angular displacements $\alpha^{s\hat{a}}_{\hat{k}}$, $\alpha^{\hat{a}\hat{b}}_{\hat{k}}$, $\alpha^{\hat{b}\hat{c}}_{\hat{k}}$ and $\alpha^{\hat{c}\hat{d}}_{\hat{k}}$. 
The adversary's goal then is to determine a combination $k'$ as an inference of $\hat{k}$ by first estimating the $\theta^{s\hat{a}}_{\hat{k}} \in \Psi_{1}$, $\theta^{\hat{a}\hat{b}}_{\hat{k}} \in \Psi_{2}$, $\theta^{\hat{b}\hat{c}}_{\hat{k}} \in \Psi_{3}$ and $\theta^{\hat{c}\hat{d}}_{\hat{k}} \in \Psi_{4}$ values from the corresponding angular displacements. In order to accomplish this, the adversary employs the linear function ($\alpha=p\theta+q$) learned earlier. Then the adversary computes $k'$ as: 
\begin{equation}
\small
\begin{aligned}
k'=\langle ((\bar{\theta}^{s\hat{a}}+s)\mod{100}), \\((-\bar{\theta}^{\hat{a}\hat{b}}+(\bar{\theta}^{s\hat{a}}+s))\mod{100}), \\ ((\bar{\theta}^{\hat{b}\hat{c}}+(-\bar{\theta}^{\hat{a}\hat{b}}+(\bar{\theta}^{s\hat{a}}+s)))\mod{100}), \\ ((-\bar{\theta}^{\hat{c}\hat{d}}+(\bar{\theta}^{\hat{b}\hat{c}}+(-\bar{\theta}^{\hat{a}\hat{b}}+(\bar{\theta}^{s\hat{a}}+s))))\mod{100})\rangle 
\end{aligned}
\label{safekeyformula}
\end{equation}

\section{Probabilistic Attack Framework}
\label{pattack}
One shortcoming of the deterministic framework is that it outputs only a \emph{single prediction}, which if incorrect, is not very useful to the adversary. A \emph{ranked list} of predictions (``\emph{close}" to the actual combination) would be useful in reducing the search space and more desirable, especially if the combination predicted by the deterministic framework is incorrect. Empirical analysis of our deterministic framework (Section \ref{eval:controlleddet}) shows that the inference error (for each inferred number in the combination) has a low standard deviation, which suggests that numbers neighboring an incorrect inference have a higher likelihood of being part of the real combination key than numbers farther away. We use this observation in the design of our probabilistic framework.

\subsection{Ranking of Padlock Key Predictions}
\label{probpadlock}
The goal of the probabilistic framework is to create an ordered list of inferred combinations, ranked based on the probability of a combination being the actual combination. We achieve this objective by giving priority to transitions closer to $\bar{\theta}^{s\hat{a}}$, $\bar{\theta}^{\hat{a}\hat{b}}$ and $\bar{\theta}^{\hat{b}\hat{c}}$ (calculated by the deterministic model), than transitions further away from it. This is done by assigning probabilities to all possible transitions in $\Phi_{1}$, $\Phi_{2}$ and $\Phi_{3}$ using three \emph{normal distributions} $\mathcal{N}(\bar{\theta}^{s\hat{a}},\sigma^2_{s\hat{a}})$, $\mathcal{N}(\bar{\theta}^{\hat{a}\hat{b}},\sigma^2_{\hat{a}\hat{b}})$ and $\mathcal{N}(\bar{\theta}^{\hat{b}\hat{c}},\sigma^2_{\hat{b}\hat{c}})$, respectively. The means and standard deviations of these distributions are learned from the deterministic model presented in Section \ref{detpadlock}. Specifically, we calculate probabilities $P(X|\alpha^{s\hat{a}}_{\hat{k}}) \sim \mathcal{N}(\bar{\theta}^{s\hat{a}},\sigma^2_{s\hat{a}})$ for all possible transitions $X\in \Phi_1$ being the actual transition performed in phase 1,  $P(Y|\alpha^{\hat{a}\hat{b}}_{\hat{k}}) \sim \mathcal{N}(\bar{\theta}^{\hat{a}\hat{b}},\sigma^2_{\hat{a}\hat{b}})$ for all possible transitions $Y\in \Phi_2$ being the actual transition performed in phase 2, and $P(Z|\alpha^{\hat{b}\hat{c}}_{\hat{k}}) \sim \mathcal{N}(\bar{\theta}^{\hat{b}\hat{c}},\sigma^2_{\hat{b}\hat{c}})$ for all possible transitions $Z\in \Phi_3$ being the actual transition performed in phase 3.

Once $P(X|\alpha^{s\hat{a}}_{\hat{k}})$, $P(Y|\alpha^{\hat{a}\hat{b}}_{\hat{k}})$ and $P(Z|\alpha^{\hat{b}\hat{c}}_{\hat{k}})$ for all possible transitions $X$, $Y$ and $Z$ are computed, the probability $P(\hat{k}=k')$ of each of the 64K possible combination keys $k'$ being the actual combination $\hat{k}$ entered by the target user can be determined as: 
\begin{equation}
\small
\begin{aligned}
P(\hat{k}=k') = P(X|\alpha^{s\hat{a}}_{\hat{k}}) P(Y|\alpha^{\hat{a}\hat{b}}_{\hat{k}}) P(Z|\alpha^{\hat{b}\hat{c}}_{\hat{k}}); \quad
\forall{(X,Y,Z)}
\end{aligned}
\label{combinedprobpadlock}
\end{equation}
Where $k'$ can be obtained by substituting $\bar{\theta}^{s\hat{a}}$, $\bar{\theta}^{\hat{a}\hat{b}}$ and $\bar{\theta}^{\hat{b}\hat{c}}$ with $X$, $Y$ and $Z$ in Equation \ref{padlockkeyformula}, respectively. All the 64K combinations $k'$ can then be \emph{ordered} or \emph{ranked} using $P(\hat{k}=k')$, with a higher value of $P(\hat{k}=k')$ indicating that $k'$ is more likely to be the actual combination $\hat{k}$. Such a ranked list of combinations, denoted as $\mathbb{\bar{K}}$, provides the adversary with a targeted search space to carry out the inference attack. If the actual combination key $\hat{k}$ lies in the top-$r$ of $\mathbb{\bar{K}}$, then the attack framework is said to succeed after $r$ attempts in the worst-case. The adversary would obviously like $r$ to be as small as possible.

\subsection{Ranking of Safe Key Predictions}
\label{probsafe}
The above probabilistic model for the padlock can be trivially extended to the safe. This is done by calculating probabilities $P(W|\alpha^{s\hat{a}}_{\hat{k}}); \forall W\in \Psi_1$, $P(X|\alpha^{\hat{a}\hat{b}}_{\hat{k}}); \forall X\in \Psi_2$, $P(Y|\alpha^{\hat{b}\hat{c}}_{\hat{k}}); \forall Y\in \Psi_3$ and $P(Z|\alpha^{\hat{c}\hat{d}}_{\hat{k}}); \forall Z\in \Psi_4$ using normal distributions $\mathcal{N}(\bar{\theta}^{s\hat{a}},\sigma^2_{s\hat{a}})$, $\mathcal{N}(\bar{\theta}^{\hat{a}\hat{b}},\sigma^2_{\hat{a}\hat{b}})$, $\mathcal{N}(\bar{\theta}^{\hat{b}\hat{c}},\sigma^2_{\hat{b}\hat{c}})$ and $\mathcal{N}(\bar{\theta}^{\hat{c}\hat{d}},\sigma^2_{\hat{c}\hat{d}})$, respectively.
Then, the probability $P(\hat{k}=k')$ of each of the $100^4$ possible combination keys $k'$ being the actual combination $\hat{k}$ entered by the target user can be determined as:
\begin{equation}
\small
\begin{aligned}
P(\hat{k}=k') = P(W|\alpha^{s\hat{a}}_{\hat{k}}) P(X|\alpha^{\hat{a}\hat{b}}_{\hat{k}}) P(Y|\alpha^{\hat{b}\hat{c}}_{\hat{k}}) P(Z|\alpha^{\hat{c}\hat{d}}_{\hat{k}})
\end{aligned}
\label{combinedprobsafe}
\end{equation}
Where $k'$ can be obtained by substituting $\bar{\theta}^{s\hat{a}}$, $\bar{\theta}^{\hat{a}\hat{b}}$, $\bar{\theta}^{\hat{b}\hat{c}}$ and $\bar{\theta}^{\hat{c}\hat{d}}$ with $W$, $X$, $Y$ and $Z$ in Equation \ref{safekeyformula}, respectively. 
All $100^4$ combinations $k'$ can then be similarly ranked in a decreasing order using $P(\hat{k}=k')$.

\subsection{Search Space Reduction}
\label{manufacturingflaws}
Although the theoretical combination space for both the Master Lock 1500T and the First Alert 2087F-BD are large enough to make manual brute-force attacks impractical, the padlock has some well-known design limitations. In practice, only a set of 4000 keys are used in the production design of Master Lock, as pointed out in a LifeHacker article \cite{lifehacker100}. Accordingly, after studying how our probabilistic attack model performs on the entire $40^3$ key space, we also analyze how our attack can improve predictions within the already reduced space of $|\mathbb{\bar{K}}|=4000$ combinations. We are not aware of similar limitations in the First Alert safe.

\section{Evaluation}
\label{evaluation}
We conduct thorough empirical evaluations of the proposed inference frameworks in order to assess their performance under realistic lock operation scenarios. Our evaluation results are outlined next.

\subsection{Experimental Setup}
We evaluate the proposed inference frameworks by means of smartwatch gyroscope data collected from a set of human subject participants who performed unlocking operations on the Master Lock 1500T padlock and the First Alert 2087F-BD safe with the watch-wearing hand. For our experiments, we employed a Samsung Gear Live smartwatch which runs Android Wear 1.5 mobile OS and is equipped with an InvenSense MP92M 9-axis Gyro + Accelerometer + Compass sensor. The smartwatch's gyroscope sensor was sampled at 200 Hz, and the samples were transmitted over a Bluetooth connection to a paired Android smartphone (specifically, a Samsung I9500 Galaxy S4). The smartphone recorded the received sensor data stream into labeled files, which were later used for training and validation (testing). All preprocessing, training and testing were performed on a server equipped with dual Intel Xeon L5640 processors and 64 GB of RAM. During the data collection, participants are clearly explained the unlocking procedure for each lock. The locks are placed on a flat table and participants sit on a chair across the table while unlocking. For the first part of our evaluation (sections \ref{eval:controlleddet} and \ref{eval:controlledprob}), we collect and use data from the participants' right hand (i.e., the right hand was used to unlock) in a controlled setting. In this setting, each combination is dictated one at a time to the participants who would then correctly enter it on the lock. Our only objective for collecting unlocking-related motion data from participants was to employ it for a realistic evaluation of the proposed inference frameworks. Our data collection procedure posed no safety or ethical risks to participants, and no private or personally identifiable information (including, combinations of personal locks/safes) was collected from participants. This study is approved by our institution's IRB.

\subsection{Deterministic Attack Framework Results}
\label{eval:controlleddet}
We evaluate the performance of the deterministic framework by measuring the standard deviations of the inferred transitions $\bar{\theta}^{ij}$ from the corresponding ground-truths $\theta^{ij}$ for each phase of the unlocking operation.
We specifically evaluated three different inference strategies: i) inferring transitions ($\textsuperscript{+}\bar{\theta}^{ij}$) solely using \emph{positive displacements} ($\textsuperscript{+}\alpha^{ij}$), ii) inferring transitions ($\textsuperscript{-}\bar{\theta}^{ij}$) solely using \emph{negative displacements} ($\textsuperscript{-}\alpha^{ij}$), and iii) \emph{averaging} inferences ($\frac{\textsuperscript{+}\bar{\theta}^{ij}+\textsuperscript{-}\bar{\theta}^{ij}}{2}$) obtained individually using positive and negative displacements. Our objective is to determine if transition inference using any one of the above displacement parameter is better than the other. 

\subsubsection{\textbf{Results for Padlock}}
\label{padlockdeteval}
The training dataset for the Master Lock 1500T padlock is composed of data collected from 3 participants (who are the authors, acting as the adversary). Each participant entered 40 different 3-digit combinations, covering all of the 120 possible transitions (40 in each of $\Phi_{1}$, $\Phi_{2}$ and $\Phi_{3}$). This data entry was repeated 3 times by each participant, resulting in a total of 9 complete datasets which is used for training the deterministic attack model. The testing dataset was collected later from a different set of 10 participants (non-authors)\footnote{The training dataset for all experiments were collected independently and before the test participants were identified/recruited, which gives us the worst-case results. However, an adversary could be more successful by personalizing the training process for the user being targeted.}. Each of these test participants entered 4 different 3-digit combinations covering 12 of the 120 possible transitions (4 in each of $\Phi_{1}$, $\Phi_{2}$ and $\Phi_{3}$), and repeated the data entry 3 times. The combination of data collected from all the 10 participants resulted in 3 complete test datasets covering all the 120 possible transitions. The data collection task is a non-trivial and time-consuming process due to the high cognitive workload associated with entering new and previously unknown combinations which resulted in a significant number of input errors by the participants. All input errors during data-collection were closely monitored and eliminated from the final datasets, and participants were asked to re-enter combinations on which errors occurred. We took utmost care to ensure that our test dataset is complete (covering all transitions) and reasonably heterogeneous (from 10 different participants) to avoid any bias in the evaluation results. The evaluation results, outlined next, are using the averaged prediction over all the 3 test datasets.

\begin{figure}[b]
\centering
\begin{subfigure}[]{0.49\linewidth}
\includegraphics[width=\textwidth]{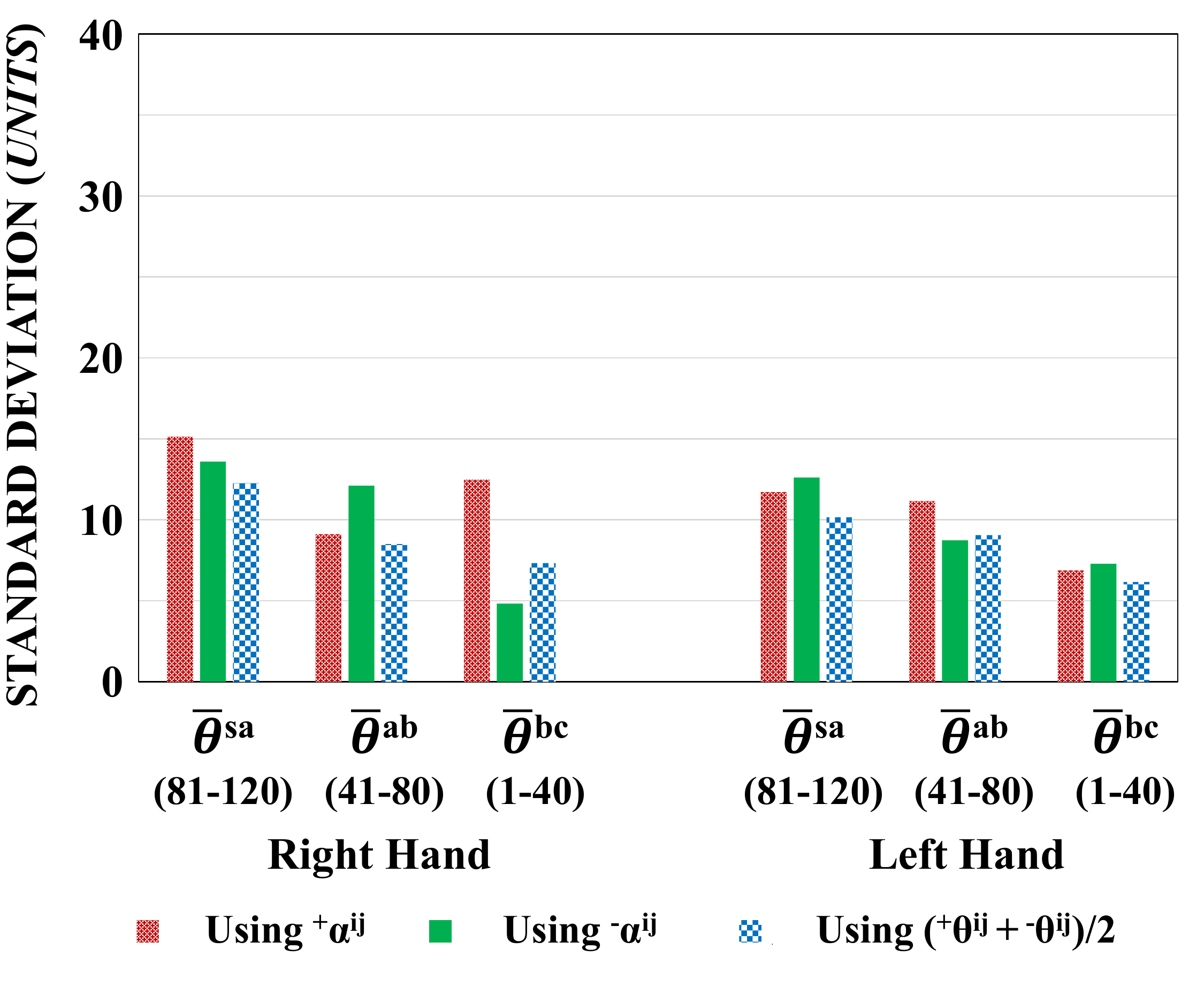}
\caption{}
\label{deviations_padlock}
\end{subfigure}
\hfill
\begin{subfigure}[]{0.49\linewidth}
\includegraphics[width=\textwidth]{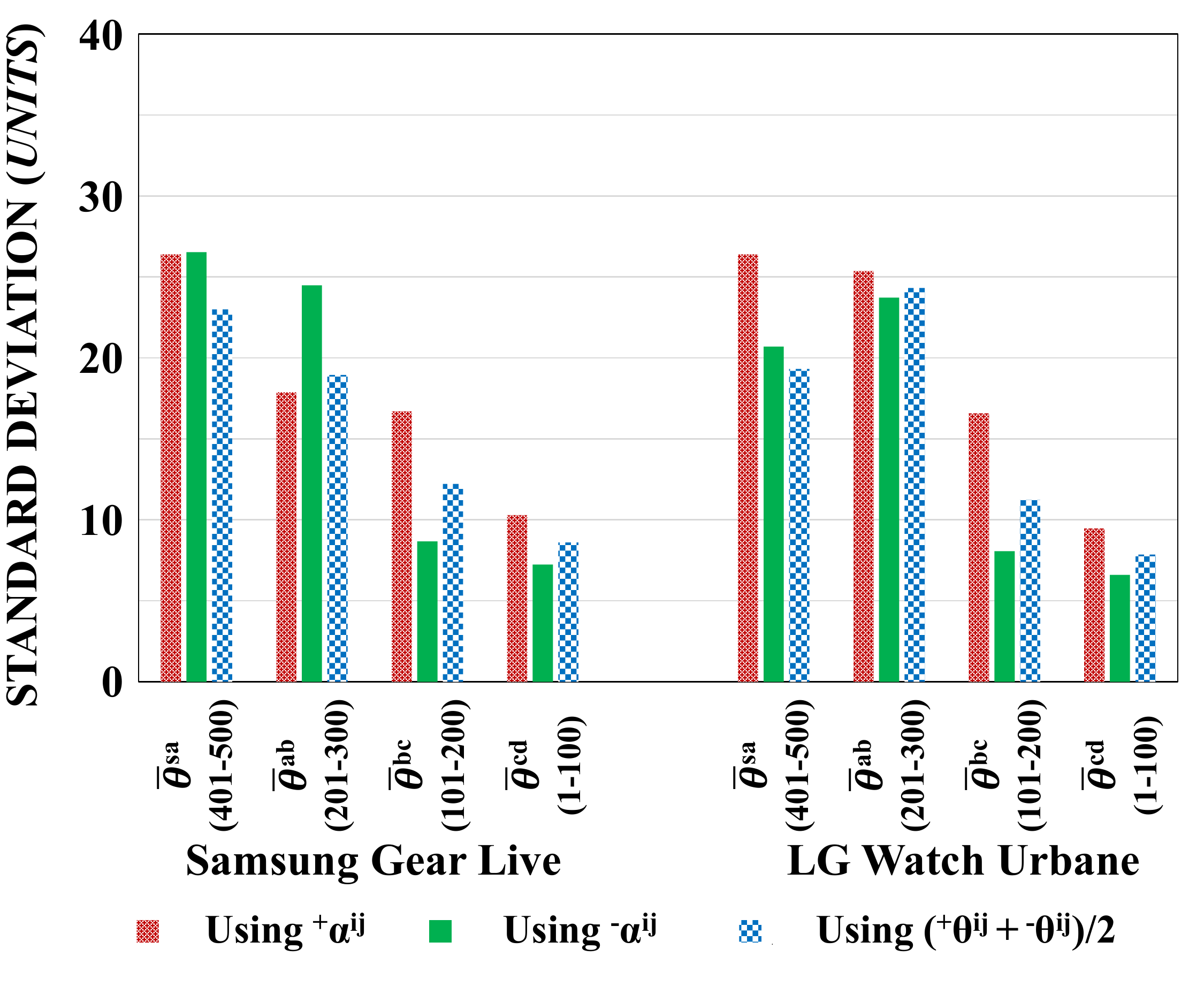}
\caption{}
\label{deviations_safe}
\end{subfigure}
\caption{Standard deviations in inference error for (a) -- three padlock phases, and (b) -- four safe phases.}
\label{deviations}
\end{figure}

Table \ref{padlocklines} shows the linear least-squares fittings for $\alpha^{sa}$, $\alpha^{ab}$ and $\alpha^{bc}$, learned from the 9 training sets. These learned linear least-squares fitting parameters ($m$ and $n$) are then used within the deterministic framework to infer the 120 unique transitions in the test dataset. 
Figure \ref{deviations_padlock} (Right Hand results) shows the standard deviations in inference errors for the inferred transitions in phase 1 ($\bar{\theta}^{sa}$), in phase 2 ($\bar{\theta}^{ab}$) and in phase 3 ($\bar{\theta}^{bc}$). We can see that the inference averaging method ($\frac{\textsuperscript{+}\bar{\theta}^{ij}+\textsuperscript{-}\bar{\theta}^{ij}}{2}$) resulted in lowest error for the inference of transitions in phase 1 (specifically, 12.27 units) and phase 2 (8.49 units), respectively. However, inference using negative displacement ($\textsuperscript{-}\alpha^{bc}$) resulted in the lowest error in phase 3 (4.82 units). \emph{We can also see that the inference of shorter transitions are more accurate than longer ones.} This observation is intuitive and could be attributed to the differences in the biomechanics of the diarthrodial joints \cite{mow2012biomechanics} of the test and training participants. These joints play an important role during the unlocking operation and the errors due to biomechanical differences could add up for longer transitions, thus making their inference more error-prone.

\begin{table}[t]
  \centering
  \scriptsize
  \caption{Linear least-squares fittings for the padlock.}
    \begin{tabular}{rcc}
    \toprule
          & $m$ (Slope) & $n$ ($\alpha$-intercept) \\
    \midrule
    $\textsuperscript{+}\alpha^{sa}$ (81-120): & 0.0836 & 0.3272 \\
    $\textsuperscript{-}\alpha^{sa}$ (81-120): & -0.1269 & 0.3714 \\
    $\textsuperscript{+}\alpha^{ab}$ (41-80): & 0.0854 & 0.9360 \\
    $\textsuperscript{-}\alpha^{ab}$ (41-80): & -0.1163 & 0.3301 \\
    $\textsuperscript{+}\alpha^{bc}$ (1-40): & 0.0737 & 2.0387 \\
    $\textsuperscript{-}\alpha^{bc}$ (1-40): & -0.1173 & 0.0061 \\
    \bottomrule
    \end{tabular}
  \label{padlocklines}
\end{table}

\begin{table}[t]
  \centering
  \scriptsize
  \caption{Linear least-squares fittings for the safe.}
    \begin{tabular}{rcc}
    \toprule
          & $p$ (Slope) & $q$ ($\alpha$-intercept) \\
    \midrule
    $\textsuperscript{+}\alpha^{sa}$ (401-500): & 0.0153 & 19.5492 \\
    $\textsuperscript{-}\alpha^{sa}$ (401-500): & -0.0266 & -8.8471 \\
    $\textsuperscript{+}\alpha^{ab}$ (201-300): & 0.0010 & 7.9046 \\
    $\textsuperscript{-}\alpha^{ab}$ (201-300): & -0.0386 & -2.3798 \\
    $\textsuperscript{+}\alpha^{bc}$ (101-200): & 0.0170 & 3.6319 \\
    $\textsuperscript{-}\alpha^{bc}$ (101-200): & -0.0460 & 0.4906 \\
    $\textsuperscript{+}\alpha^{cd}$(1-100): & 0.0305 & 1.7663 \\
    $\textsuperscript{-}\alpha^{cd}$ (1-100): & -0.0483 & -0.1058 \\
    \bottomrule
    \end{tabular}
  \label{safelines}
\end{table}

\subsubsection{\textbf{Results for Safe}}
\label{safedeteval}
The training dataset for the First Alert 2087F-BD safe is composed of data collected from 3 participants (who are the authors). Each participant entered 100 different 4-digit combinations, covering all of the 400 possible transitions (100 in each of $\Psi_{1}$, $\Psi_{2}$, $\Psi_{3}$ and $\Psi_{4}$), which resulted in 3 complete training datasets. Testing dataset was collected later from a set of 10 different participants (non-authors), where each participant entered 2 different 4-digit combinations covering 8 of the 400 possible transitions (2 in each of the transition sets $\{405, 410, 415, ... 500\}$, $\{205, 210, 215, ... 300\}$, $\{105, 110, 115, ... 200\}$ and $\{5, 10, 15, ... 100\}$). Each participant repeated entering each combination 3 times, which resulted in 3 partially complete test datasets of 80 evenly distributed transitions. Due to a slightly more complex and longer unlocking procedure of the safe (compared to the padlock), we observed a larger number of participant errors during combination entry. As before, all input errors were closely monitored and removed from the final datasets. Due to a large combination space, in addition to the more complex unlocking procedure, we restricted ourselves to only partial test datasets for the safe. However, we made sure that the test dataset is uniform in terms of the distribution of the various transitions and the participants that recorded those transitions to avoid any bias in the evaluation results. The evaluation results, outlined next, are using the averaged prediction over all the 3 test datasets.

Table \ref{safelines} shows the linear least-squares fittings for $\alpha^{sa}$, $\alpha^{ab}$, $\alpha^{bc}$ and $\alpha^{cd}$, learned from the 3 training sets. These learned linear least-squares fitting parameters ($p$ and $q$) are then used to infer the 80 unique transitions in the test datasets. The standard deviations in inference errors for the inferred transitions in phase 1 ($\bar{\theta}^{sa}$), in phase 2 ($\bar{\theta}^{ab}$), in phase 3 ($\bar{\theta}^{bc}$) and in phase 4 ($\bar{\theta}^{cd}$) are outlined in Figure \ref{deviations_safe} (Samsung Gear Live results). We can see that the inference averaging method resulted in the lowest error for the inference of transitions in phase 1  (specifically, 22.99 units), while inference using positive displacement ($\textsuperscript{+}\alpha^{ab}$) resulted in the lowest error for the inference of transitions in phase 2 (17.86 units). For transitions in phase 3 and phase 4, inference using the corresponding negative displacements (i.e., $\textsuperscript{-}\alpha^{bc}$ and $\textsuperscript{-}\alpha^{cd}$) resulted in lowest errors (8.66 and 7.23 units, respectively). \emph{Similar to the padlock case, we can observe that inference of shorter transitions in safe combinations are more accurate. Moreover, we also observe that the standard deviations of inference errors for the safe are relatively higher compared to the padlock.} We believe that this is due to the higher concentration of numbers on the safe's lock dial, compared to the padlock's dial, for the same angular displacement.

\subsection{Probabilistic Attack Framework Results}
\label{eval:controlledprob}
We evaluate the performance of the probabilistic attack model by evaluating the overall success probability of test combination keys being present in the top-$r$ of their corresponding ranked inferred combination sets.

\begin{figure*}[b]
\centering
\begin{subfigure}[]{0.32\linewidth}
\includegraphics[width=\linewidth]{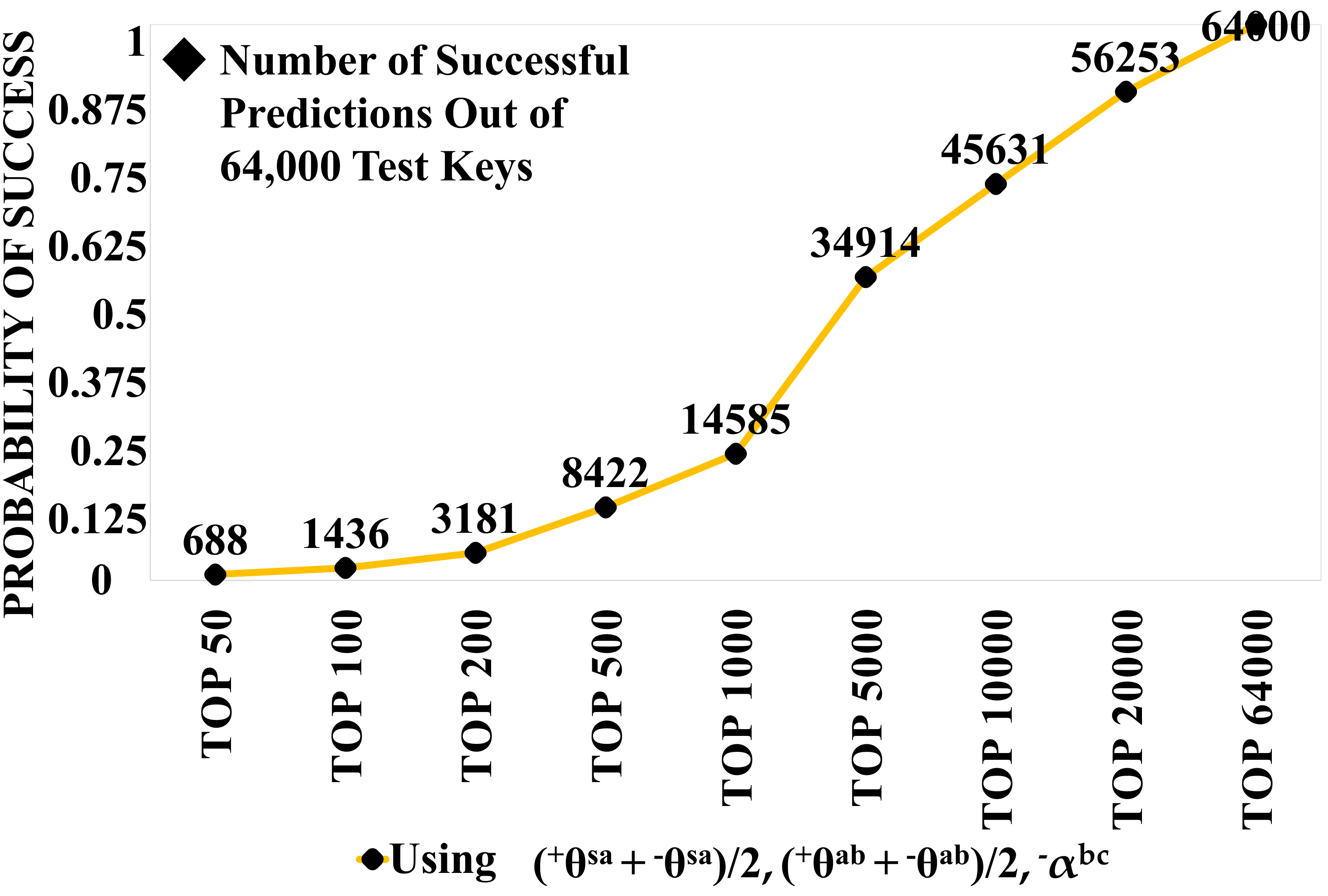}
\caption{}
\label{padlockproball}
\end{subfigure}
\hfill
\begin{subfigure}[]{0.32\linewidth}
\includegraphics[width=\linewidth]{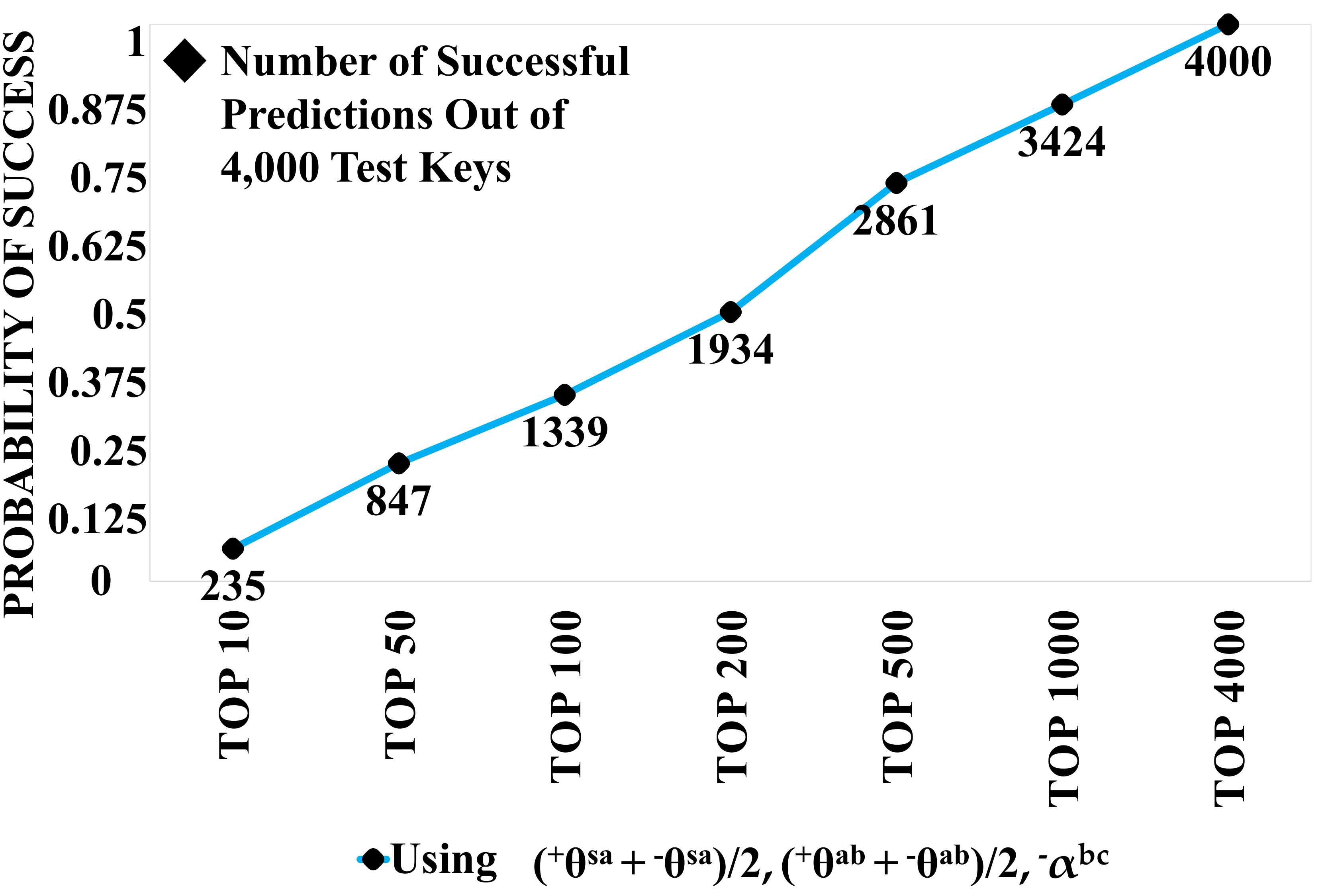}
\caption{}
\label{padlockprob4000}
\end{subfigure}
\hfill
\begin{subfigure}[]{0.32\linewidth}
\includegraphics[width=\linewidth]{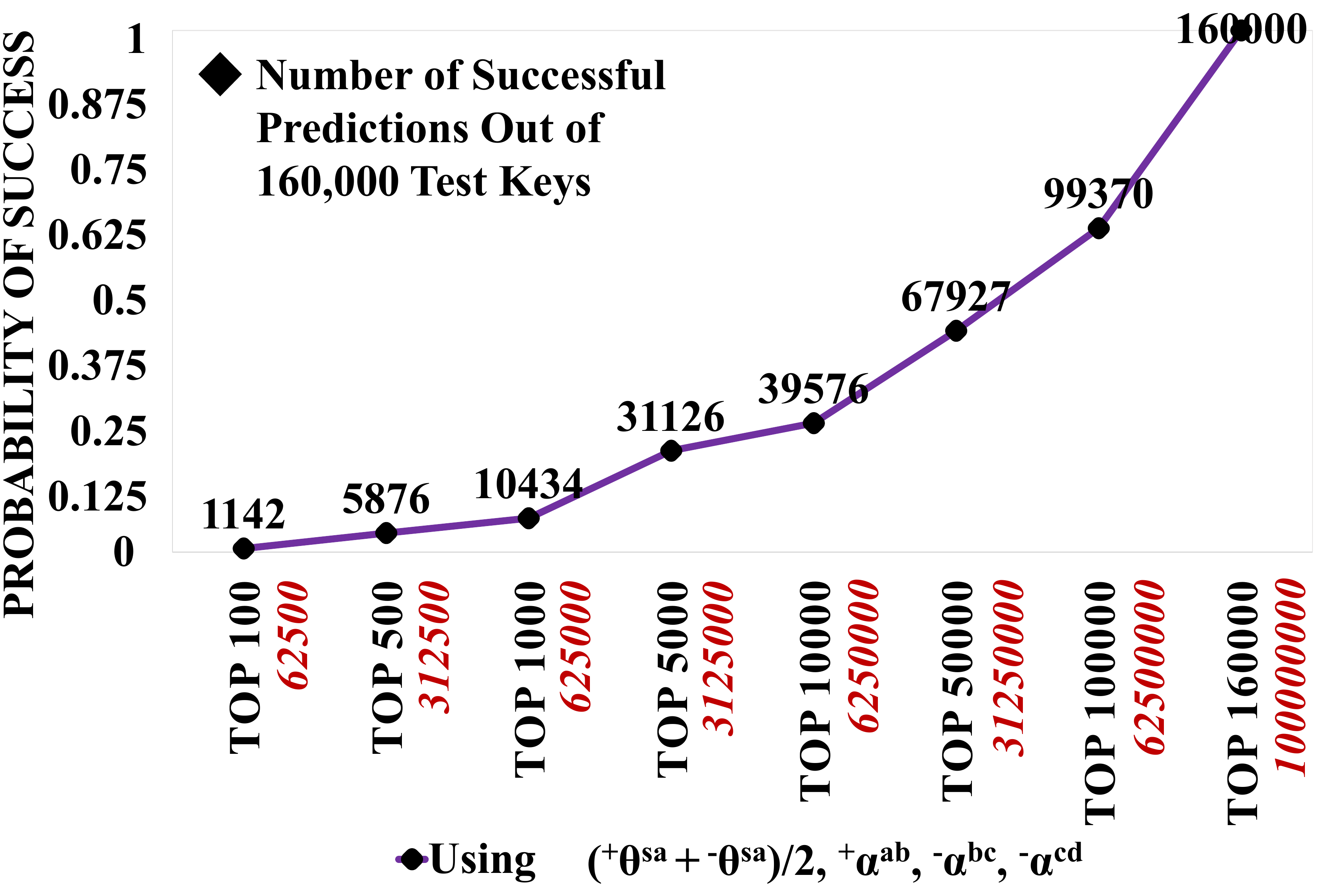}
\caption{}
\label{safeproball}
\end{subfigure}

\begin{subfigure}[]{0.32\linewidth}
\includegraphics[width=\linewidth]{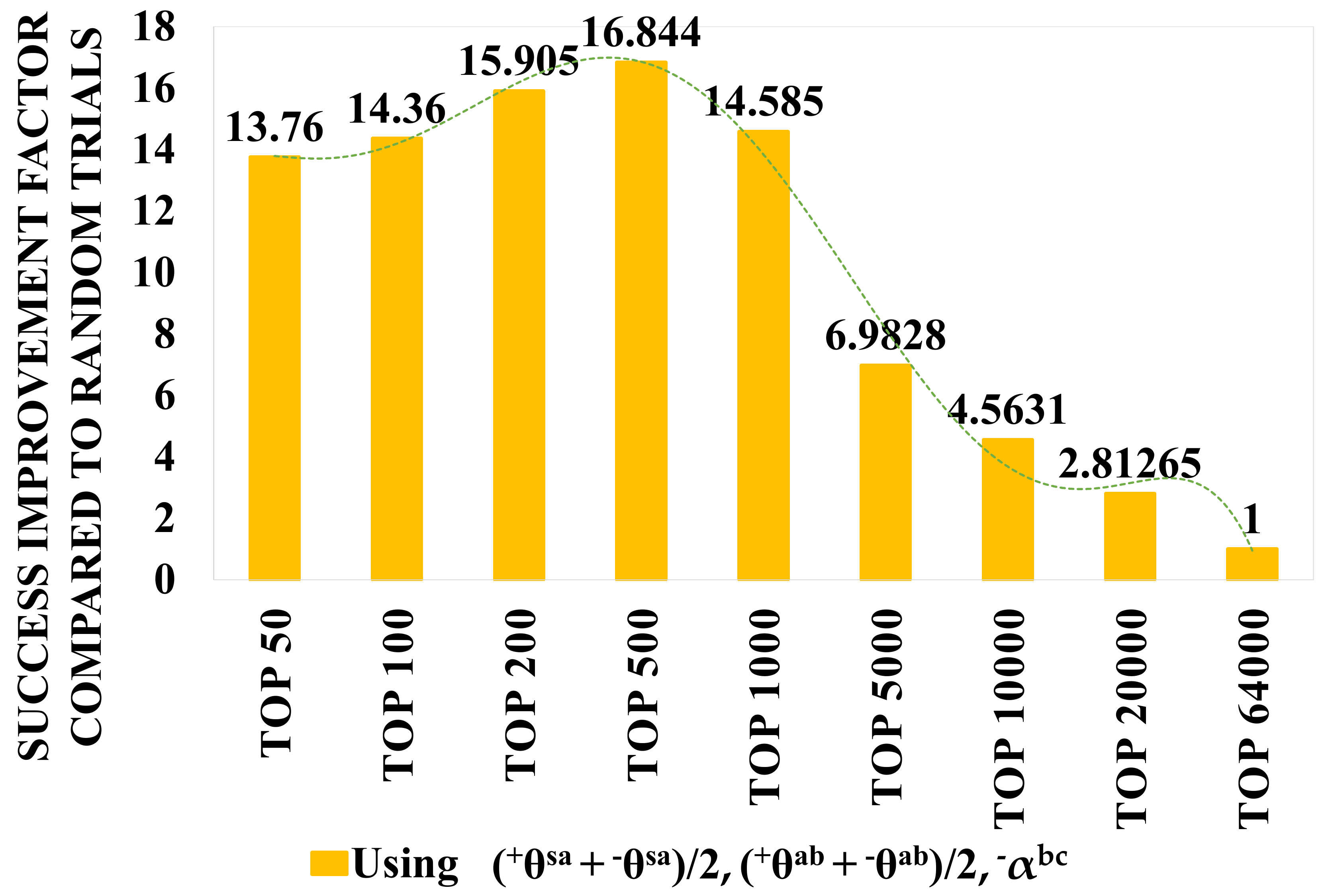}
\caption{}
\label{padlockproballfactor}
\end{subfigure}
\hfill
\begin{subfigure}[]{0.32\linewidth}
\includegraphics[width=\linewidth]{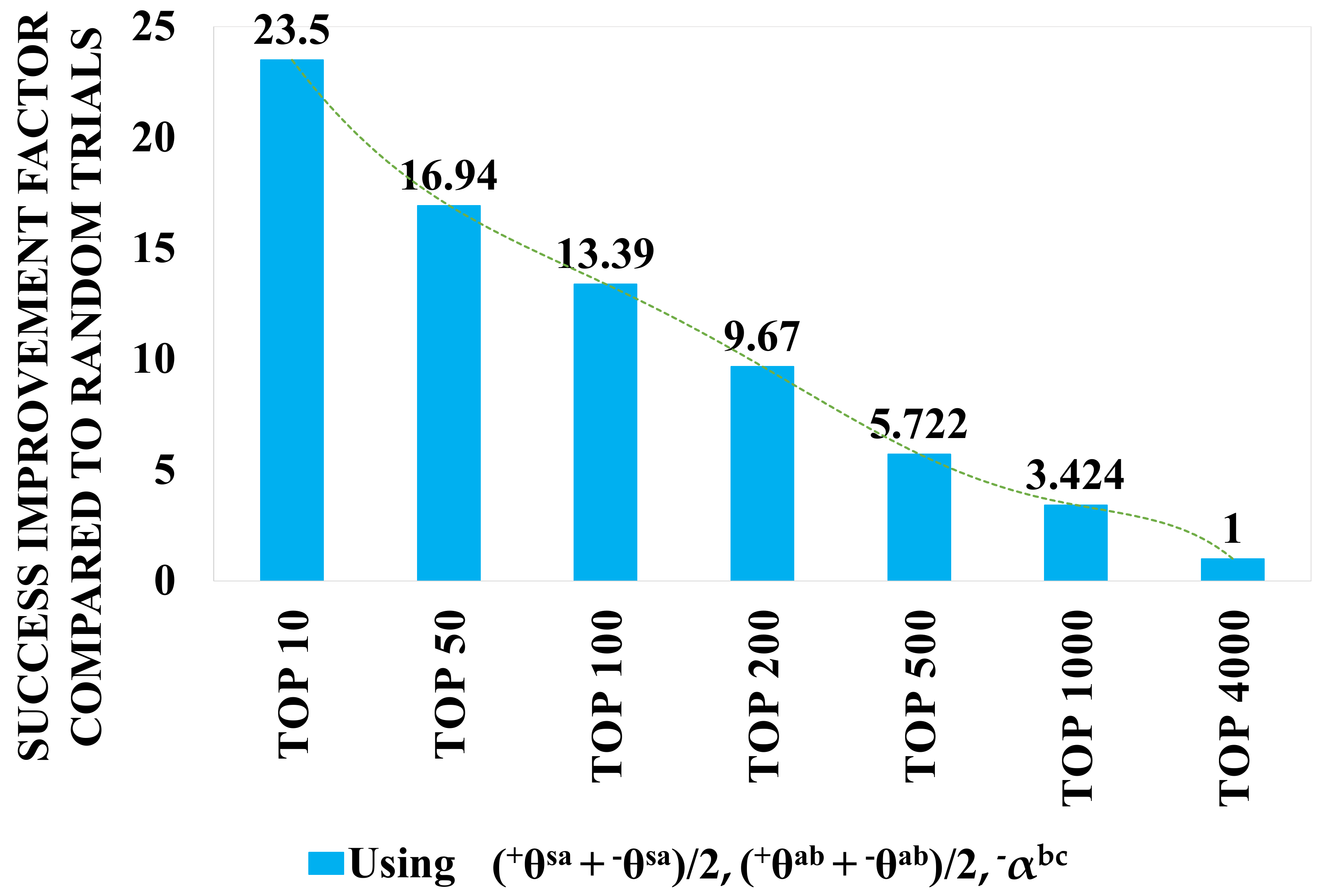}
\caption{}
\label{padlockprob4000factor}
\end{subfigure}
\hfill
\begin{subfigure}[]{0.32\linewidth}
\includegraphics[width=\linewidth]{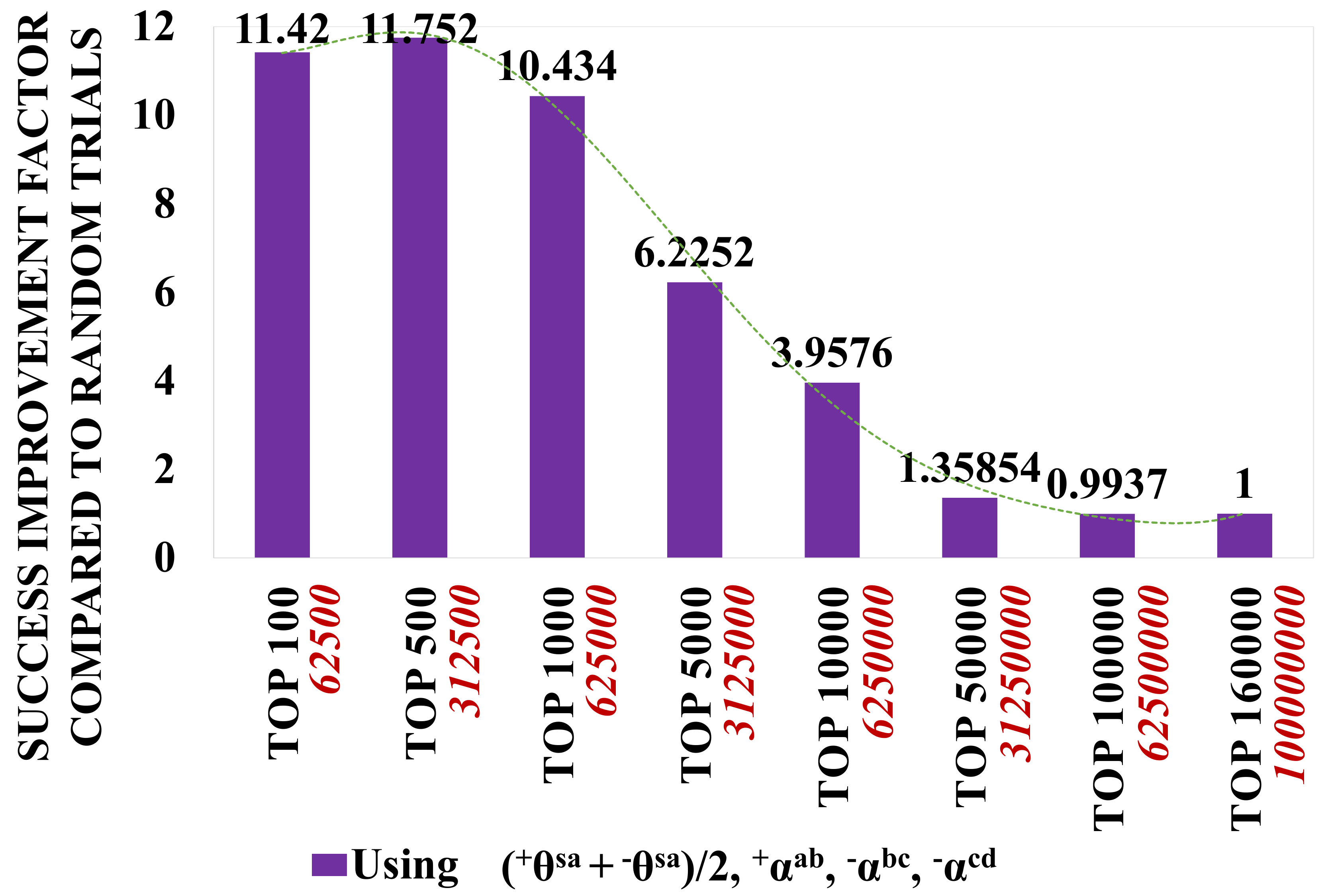}
\caption{}
\label{safeproballfactor}
\end{subfigure}
\caption{(a) $-$ Top-$r$ success probabilities for inferred padlock combinations using 64K test combinations; (b) $-$ Top-$r$ success probabilities for inferred padlock combinations using 4K test combinations; (c) $-$ Top-$r$ success probabilities for inferred safe combinations using 160K test combinations; (d), (e), (f) $-$ Success improvement factors compared to random trials, for the padlock test set of 64K test combinations, padlock test set of 4K test combinations and safe test set of 160K test combinations, respectively.}
\label{keypredictions}
\end{figure*}

\subsubsection{\textbf{Padlock Key Predictions (64K)}}
\label{padlockkeypredictions}
We first evaluate the success probability of finding an entire padlock test combination key within the top-$r$ of the corresponding set of 64K candidate keys, ranked using the probabilistic model. The 64K unique test combinations were obtained by \emph{combination} of $\Phi_{1}$, $\Phi_{2}$ and $\Phi_{3}$ datasets. In this case, rather than using all the three methods for the inference of the individual transitions of the test combination, i.e., inference using only positive displacements, only negative displacements, and averaging individual inferences, we optimize the overall combination inference by selecting the inference method with the lowest error in each phase, for inferring transitions of the test combination key in that phase. Thus for all the 64K test padlock combinations, the first two phases were inferred using inference averaging method, and the third phase using negative displacements. 
The value of $r$ was increased from 50 to 64,000 in varying steps. Figure \ref{padlockproball} shows the success probability of finding a test combination within the top-$r$ of the corresponding probabilistically ranked (using Equation \ref{combinedprobpadlock}) set of candidate keys. 688 test combinations (out of a total of 64K test combinations) were found in the top-50 of their corresponding ranked inferred combination set, which equates to a 1.07\% overall probability of success. Compared to this, the probability of correctly picking a test combination after 50 random guesses is only 0.078\%. This implies that for $r=50$ the proposed probabilistic model achieves an improvement by a factor of 13.76 over random guessing. Despite the low overall success probability, the above results are indicative of the fact that certain combinations (688 test combinations) are easier to infer than others. Figure \ref{padlockproballfactor} show similar improvements factors for all other top-$r$ cases. These results indicate that an adversary can significantly reduce the search space, and still have high probability of success. As a result, \emph{the cumulative probability of success using the probabilistic model is much higher with `limited' number of trials, compared to random guessing or the deterministic attack.}

\subsubsection{\textbf{Padlock Key Predictions (4K)}}
\label{padlockkeypredictions4k}

We again evaluate the success probability of finding an entire test combination key within the top-$r$ of the corresponding set of candidate keys ranked using the probabilistic model, but this time using the only the 4K implemented padlock combinations (outlined in Section \ref{manufacturingflaws}) as test combinations. Similar to the 64K analysis, for all the 4K test padlock combinations, the first two phases were inferred using inference averaging method, and the third phase using negative displacements. The value of $r$ was increased from 10 to 4,000 in varying steps. Figure \ref{padlockprob4000} shows the success probability of finding a test combination within the top-$r$ of the corresponding probabilistically ranked (using Equation \ref{combinedprobpadlock}) set of candidate keys. 235 test combinations (out of a total of 4K test combinations) were found in the top-10 of their corresponding ranked inferred combination set, which equates to a 5.87\% overall probability of success. Compared to this, the probability of correctly picking a test combination (among all the implemented keys) after 10 random guesses is only 0.25\%. This implies that for $r=10$ the proposed probabilistic model achieves an improvement by a factor of 23.5 over random guessing. Figure \ref{padlockprob4000factor} show similar improvements factors for all other top-$r$ cases. \emph{These results indicate that an adversary can significantly reduce the combination search space by leveraging on both known mechanical flaws and eavesdropped wrist-movements.}

\subsubsection{\textbf{Safe Key Predictions (160K)}}
\label{safekeypredictions}
We next evaluate the success probability of finding an entire test combination key within the top-$r$ of the corresponding set of candidate keys ranked using the probabilistic model. For this analysis, we test 160K safe combinations $k=\langle a,b,c,d\rangle $ distributed evenly across the entire $100^4$ combination space ($\theta^{s\hat{a}} \in \{405, 410, 415, ... 500\}$, $\theta^{\hat{a}\hat{b}} \in \{205, 210, 215, ... 300\}$, $\theta^{\hat{a}\hat{b}} \in \{105, 110, 115, ... 200\}$ and $\theta^{\hat{a}\hat{b}} \in \{5, 10, 15, ... 100\}$). 
The 160K unique test combinations were obtained by \emph{combination} of $\Psi_{1}$, $\Psi_{2}$, $\Psi_{3}$ and $\Psi_{4}$ datasets.
Similar to padlock key predictions, we optimize the overall combination inference by selecting the inference method with the lowest error in each phase, for inferring transitions of the test combination key in that phase. Thus for all the 160K test safe combinations, the first phase was inferred using inference averaging method, the second phase was inferred using positive displacements, and the last two phases using negative displacements. 
The value of $r$ was increased from 100 to 160,000 in varying steps. It should be noted that in the case of the safe, we only probabilistically rank the (evenly distributed) 160K keys appearing in the test set rather than the entire safe combination space of $100^4$. This is primarily due to the computational challenge associated with computing probabilities for, and then ranking, 100 million combination keys for each of the 160K test combinations, which is an extremely time-consuming process. The adversary, however, does not have a similar problem because the adversary has to rank the entire combination space of $100^4$ for only a few test keys, which is relatively easier to compute. Figure \ref{safeproball} shows the success probability of finding a test combination within the top-$r$ of the corresponding probabilistically ranked set of candidate keys. 5876 test combinations (out of a total of 160K test combinations) were found in the top-500 of their corresponding ranked inferred combination set, which equates to a 3.67\% overall probability of success. Compared to this, the probability of correctly picking a test combination after 500 random guesses is only 0.31\%. This implies that for $r=500$ the proposed probabilistic model achieves an improvement by a factor of 11.42 over random guessing. A straightforward extrapolation of $r$ (multiplying it with a factor of $5^4$) puts the value of $r$ at $312500$ for achieving similar improvement if the entire combination space of $100^4$ combinations is ranked. Readers should note that labels marked in red in Figure \ref{safeproball} are extrapolated values of $r$.
Figure \ref{safeproballfactor} show similar improvements factors for all other top-$r$ cases. \emph{These results indicate that the proposed framework can achieve significant reduction of the combination search space for the safe as well.}

\subsection{Cross-Device Performance}
\label{eval:controlledcrossdevice}
So far we have evaluated our inference frameworks in a same-device setting where the same smartwatch hardware (Samsung Gear Live smartwatch with an InvenSense MP92M sensor) was used for collecting both the training and testing datasets. However in a practical setting, an adversary may be unaware of, or may not possess, the precise wrist-wearable hardware used by the target user. Thus, it is critical to assess the performance of the inference frameworks when different wrist-wearable hardwares are used for training and testing (attack) purposes. In other words, a comparison of the earlier evaluation results with results using test data from a different smartwatch would tell us if the proposed inference frameworks are inter-operable across different devices. For brevity, we analyze the cross-device performance of the inference frameworks only for the First Alert 2087F-BD safe. 
For this, we collect the same set of test data for the safe as detailed in Section \ref{safedeteval} by using a LG Watch Urbane smartwatch equipped with an on-board InvenSense M651 6-axis Gyro + Accelerometer sensor (sampled at 200 Hz) and running Android 2.0 mobile OS. We then employ the linear function $\alpha=p\theta+q$ (Table \ref{safelines}), trained from the data collected with a Samsung Gear Live (as outlined in section \ref{safedeteval}). 

A comparison of the standard deviations in inference error (Figure \ref{deviations_safe}) does not show significant change in prediction results we observed earlier. A \emph{pair-wise two-tailed $t$-test} \cite{zimmerman1997teacher} of all the values in both set of results, resulted in \emph{$t=0.11;p=0.915$. The small value of $t$ indicates that there exists minor difference between the two sets of results. However, due to the high $p$ value (which implies that our results have likely occurred by chance), we cannot conclusively say that an adversary can train the inference models using data from one device and use these trained models to carry out inference attacks on data from a different wrist-wearable device.} That being said, it is not difficult for an adversary to train a new model according to the target user's wrist-wearable device.

\subsection{Cross-Hand Performance}
\label{eval:controlledcrosshand}
All evaluations of our inference models so far have been accomplished using training and testing datasets collected from subjects who only used their right hand for the unlocking operation.
However in a practical setting, a target user may not perform the unlocking operation with the same hand that the adversary has trained its models on. Thus, it is important to assess the performance of the proposed inference frameworks when training and testing data corresponding to the unlocking operation comes from different hands. In other words, we would like to analyze if the proposed inference models trained using unlocking data from one hand (say, right) can be used to infer combinations entered using the other hand (say, left). For this we collect the same test data for the padlock as detailed in Section \ref{padlockdeteval}, but this time the 10 participants wore the Samsung Gear Live smartwatch on their left hand and entered the test combinations on the padlock with their left hand. We then employ the linear function $\alpha=m\theta+n$, trained earlier using the right hand data (Table \ref{padlocklines}), to infer transitions in each phase using the deterministic model.

A comparison of the standard deviations in inference error (Figure \ref{deviations_padlock}) does not show significant change in prediction results we observed earlier using same-hand predictions. \emph{A pair-wise two-tailed $t$-test of all the values in both set of results, resulted in $t=1.33;p=0.219$. The small value of $t$ along with a low $p$ value indicates that the mean difference between the two sets of results is not significant, with a low probability that our results occurred by chance. We can therefore conclude that an adversary can focus on training a single model with either hand's data, and use it on both left and right handed targets.} Nevertheless, it is not difficult for an adversary to train two different models, one per hand.

\subsection{Real-Life Detection and Prediction}
\label{eval:uncontrolled}
Next, we evaluate the performance of our unlocking activity recognition algorithm (Section \ref{activityrecognition}) and inference framework under a \emph{real-life setting}. To facilitate a real-life experiment with three new participants, we handed out a Samsung Gear Live smartwatch, a paired smartphone and a padlock, for them to take home. The watch was installed with our recording application and the unlock activity recognition algorithm. We collected $x$-axis gyroscope data for the duration of approximately 1 day, during which the participants were instructed to perform at least three padlock unlock operations with a 3-digit combination of their choice (among the 4K mechanically valid combinations), at random intervals. \emph{Overall, our unlock activity recognition algorithm yielded 100\% recall and 80\% precision, with a total of 12 true positives, 3 false positives and 0 false negative}. Interestingly, the 3 false positives were reportedly due to activities similar to padlock unlocking, such as when washing hands after rotating a washer tap/faucet, and while using a screw driver. Next, we evaluate the prediction accuracy of the secret combination entered by each participant by using the last three instance of their unlocking gyroscope time-series data, as extracted from the entire day's data. Applying the same inference model for ranking keys among the 4K implement keys, used in Section \ref{padlockkeypredictions4k}, the real key entered by the three participants were ranked at 42, 85 and 112 (out of 4000). This demonstrates the extent to which the proposed attacks can reduce the combination search space even in uncontrolled real-life settings.

\section{Discussions}
\label{discussions}
\subsection{Characteristics of Inferred Combinations}
Results of our deterministic attack model indicated that shorter transitions can be more accurately inferred than longer transitions. To see if this phenomenon carries over to full combinations as well, we analyzed the length (in terms of transition units) of the 235 padlock combinations (out of 4K) that were successfully inferred within top-10 trials (Figure \ref{padlockprob4000}). The shortest padlock combination can be of 123 transition units $(81+41+1)$, where as the longest combination can be of 240 transition units $(120+80+40)$. On this scale, 91.06\% of the 235 padlock combinations that were successfully inferred within top-10 trials, were shorter than 150 transition units. Based on this observation, we can conclude that \emph{key combinations that require less rotational displacement have better inference probability, and users should avoid purchasing locks preset with such combinations.}

\subsection{Limitations}
\noindent
\textbf{Starting Point:}
Without a known starting point the adversary will have to try all numbers on the dial as the starting point, thereby significantly increasing the average number of trials it would take to be successful. However, because the adversary will require physical access to the lock in order to try predicted keys, it is not unrealistic to assume that they can also learn or preset the starting point. The starting point can be any number on the dial; only the key inference functions (Equations \ref{padlockkeyformula} and \ref{safekeyformula}) must be appropriately initialized according to that starting point. Moreover, the learned least-squares (Tables \ref{padlocklines} and \ref{safelines}) are not affected by a change in the starting point during the attack phase.

\noindent
\textbf{Kinesiological Factors:}
Factors such as the grasping style, hand size, and muscle strength have a significant effect on the biomechanics of the diarthrodial joints of a target user performing an unlocking operation. While we did not encounter any participant in our study with significantly different unlocking styles, it is possible for an adversary to come across a target whose unlocking style is significantly different. However, a competent adversary may be able to train a variety of models based on different kinesiological factors, and use an appropriate model for each target. Further user study is required to understand if unlocking styles can be classified in to characteristically unique groups.

\noindent
\textbf{Affected Users:}
Our attack assumes that the user wears his/her wrist-wearable on the hand used to unlock the padlock or safe. This may not always be the case, causing the attack to fail. While we did not find any statistics in the literature to deduce the percentage of users who use the same hand for both, according to an on-going online poll with about 5000 participants \cite{ablogtowatchpoll}, approximately 38.23\% of users prefer to wear watches on their dominant hand. Assuming most users use their dominant hand for unlocking padlocks and safes, a significant number of users can be affected by the proposed attack. Moreover, with the advent of fitness trackers (most of which also have gyroscope sensors), users tend to wear their watches and wrist-based fitness trackers on different hands. Regardless of the exact statistics, we hope that this work will create awareness of this threat.

\begin{table}[b]
  \small
  \centering
  \caption{Popular padlocks and safes retailed by Amazon and Walmart.}
    \begin{tabular}{A{1.8cm} A{1.8cm} A{1.2cm} A{1.8cm}}
    \toprule
    Product & Combination Length $\downarrow$ & Numbers on Dial & Mechanical Limitations \\
    \midrule \midrule
    Master Lock 1500iD Speed Dial & Unlimited & 4     & 7501 unique states \cite{huebler2009} \\ \midrule
    First Alert 2087F-BD Safe & 4     & 100   & Unknown \\ \midrule
    SentrySafe SFW082CTB & 3     & 100   & Unknown \\ \midrule
    Master Lock 1500T & 3     & 40    & 4000 used combinations \\ \midrule
    Master Lock Padlock 1588D & 3     & 20    & Unknown \\
    \bottomrule
    \end{tabular}
  \label{rotarylocklist}
\end{table}

\noindent
\textbf{Generalizability:} The proposed attack frameworks can be easily extended to work with any other rotation-based mechanical combination lock, with different length of combination keys and different sequence of key entry directions (clockwise/counter-clockwise). But, according to the trends in our evaluation results, it can be challenging to infer longer combination keys (5 or more numbers) and on locks with more concentrated dials (more numbers on the dial face), with high accuracy. However, in a brief study of the most popular consumer-grade padlocks and safes, we found that 5 or more number combination padlocks and safes are very rare in the retail market (Table \ref{rotarylocklist}). There are several consumer-grade padlocks and safes available for purchase, from different manufacturers and retailers. However, most of them employ similar, if not identical, working mechanisms. In Table \ref{rotarylocklist}, we list the most popular products with rotation-based locks, which are representative of its type. Similar locks produced by different manufacturers  are faced with the same level of threat, unless the manufacture introduces additional design changes. For example, a Hollon Home Safe 310D and the First Alert 2087F-BD Safe, both have 4-number combinations with 100 numbers on the dial, resulting in $100^4$ possible combinations. However, many larger enterprise-grade safes and vaults are equipped with larger dials which can translate in to more perceptible gyroscope readings on the wrist. As a result, a more accurate inference may be possible for locks with significantly larger dials. The principles used in our attack models can also be used to infer private activities pertaining to other forms of rotary wrist movements, such as numbers entered on a rotary telephone dial, driving trajectory on a steering wheel, etc.

\subsection{Mitigations}
Users can take few preventive measures to avoid falling victim to the proposed attack. A simple measure could be to use the hand without any wrist-wearable for unlocking, or to take off any wrist-wearables before unlocking. Users could also inject noise in the data by shaking their hand in between the unlock operation. More complex protection mechanisms can include dynamic access control of zero-permission sensors such as the gyroscope. As some of the previous works suggested \cite{cappos2014blursense,maiti2016smartwatch}, a dynamic access control can take advantage of contextual information to automatically cut-off sensor access when users are detected to be vulnerable. A potential solution in this direction could be to use our unlocking activity recognition algorithm (presented in Section \ref{activityrecognition}) in a real-time fashion, so as to disable the gyroscope after the first few \textit{spins}.

\subsection{Other Attack Vectors}
Padlock and safe combinations are also susceptible to other forms of \emph{non-intrusive} attacks, such as visual shoulder-surfing when the target user is unlocking. Visual access to a lock's dial when the user is unlocking, can result in more precise key inference than our wrist motion based inference framework. However, there is high likelihood that the user will notice a visual observer (human or camera), and shield their unlocking activity. It may also be possible to use visual on the user (instead of the lock) to perform timing based inference attacks \cite{foo2010timing}.

\section{Conclusions}
\label{conclusion}

In this paper, we presented a new motion-based attack to infer mechanical lock combinations from smartwatch gyroscope data. A comprehensive evaluation using a commercial padlock and safe demonstrated that our framework can significantly reduce the combination search space for an adversary. The combination key search space can be further reduced in case of the padlock by leveraging on mechanical design flaws. We also observed that the performance of the proposed inference frameworks do not significantly degrade when model training and inference tasks are carried out using different smartwatch hardwares or different unlocking hands. Finally, we also demonstrated the efficacy of the proposed inference attack in a real-life setting.

\section*{Acknowledgment}
Research reported in this publication was supported by the Division of Computer and Network Systems (CNS) of the National Science Foundation (NSF) under award number 1828071 (originally 1523960).

\bibliographystyle{abbrv}
\bibliography{sigproc}

\section*{Appendix}
In an earlier version of this paper, we made an incorrect conclusion of the statistical significance of our results presented in Sections \ref{eval:controlledcrossdevice} and \ref{eval:controlledcrosshand}, related to cross-device and cross-hand application of our attack models. A low $p$ value indicates that the result ($t$ value) obtained is not by `chance', and vice versa. This version of the paper rectifies the conclusions made based on the observed $p$ values in Sections \ref{eval:controlledcrossdevice} and \ref{eval:controlledcrosshand}.

\end{document}